\documentclass[11pt,a4paper]{article}
\usepackage{authblk}
\usepackage[english]{babel}
\usepackage[T1]{fontenc}
\usepackage[utf8]{inputenc}
\usepackage{booktabs}
\usepackage{xcolor}
\usepackage[bottom]{footmisc}
\usepackage[titletoc]{appendix}
\usepackage{lscape}
\usepackage{textcomp}
\usepackage{eurosym}
\usepackage{multirow}
\usepackage[autolanguage]{numprint}
\usepackage[a4paper,top=3cm,bottom=2cm,left=3cm,right=3cm,marginparwidth=1.75cm]{geometry}
\usepackage{amsmath,amssymb,amsthm,bbm}
\usepackage{mathtools}
\usepackage{listings}
\usepackage[ruled,vlined]{algorithm2e}
\input{alphabet}
\input{abrege}
\newsavebox{\fminibox}
\newlength{\fminilength}
\newenvironment{fminipage}[1][\linewidth]
  {\setlength{\fminilength}{#1}
   \begin{lrbox}{\fminibox}\begin{minipage}{\fminilength}}
  {\end{minipage}\end{lrbox}\noindent\fbox{\usebox{\fminibox}}}

  \def\+{^\dagger}

\def\nequiv{\not\kern-.05em\equiv}
\def\egal{\kern-.5em=\kern-.5em}        
\def\propt{\kern-.2em\propto\kern-.2em} 

\def\intdouble{\int\kern-0.3em\int}
\def\inttriple{\int\kern-0.3em\int\kern-0.3em\int}

\def\rond#1{\overset{\kern-0.33em~_\circ}{#1}}
\def\rondit[#1]#2{\overset{\kern#1~_\circ}{#2}}

\def\babs{\begin{abstract}}             \def\eabs{\end{abstract}}
\def\barr{\begin{array}}                \def\earr{\end{array}}
\def\bcc{\begin{center}}                \def\ecc{\end{center}}

\def\bdes{\begin{description}}          \def\edes{\end{description}}
\def\bdoc{\begin{document}}             \def\edoc{\end{document}}
\def\ben{\begin{enumerate}}             \def\een{\end{enumerate}}
\def\beqn{\begin{eqnarray}}             \def\eeqn{\end{eqnarray}}
\def\beqnl#1{\beqn\label{#1}}           \def\eeqnl#1{\label{#1}\eeqn}
\def\beqnx{\begin{eqnarray*}}           \def\eeqnx{\end{eqnarray*}}
\def\bseqn{\begin{subeqnarray}}         \def\eseqn{\end{subeqnarray}}
\def\beq#1\eeq{\begin{equation}#1\end{equation}}
\def\bal#1\eal{\begin{align}#1\end{align}}
\def\balx#1\ealx{\begin{align*}#1\end{align*}}
\def\beqx{$$}                           \def\eeqx{$$}
\def\bfig{\protect\begin{figure}}       \def\efig{\protect\end{figure}}
\def\bfigx{\protect\begin{figure*}}     \def\efigx{\protect\end{figure*}}
\def\bfigt{\protect\begin{figurette}}   \def\efigt{\protect\end{figurette}}
\def\bfl{\begin{flushleft}}             \def\efl{\end{flushleft}}
\def\bfr{\begin{flushright}}            \def\efr{\end{flushright}}
\def\bit{\begin{itemize}}               \def\eit{\end{itemize}}
\def\bmi{\begin{minipage}}              \def\emi{\end{minipage}}
\def\bfmi{\begin{fminipage}}            \def\efmi{\end{fminipage}}
\def\bpic{\begin{picture}}              \def\epic{\end{picture}}
\def\bqu{\begin{quote}}                 \def\equ{\end{quote}}
\def\bqun{\begin{quotation}}            \def\equn{\end{quotation}}
\def\bsl{\begin{slide}}                 \def\esl{\end{slide}}
\def\btabb{\begin{tabbing}}             \def\etabb{\end{tabbing}}
\def\btabl{\begin{table}}               \def\etabl{\end{table}}
\def\btablx{\begin{table*}}             \def\etablx{\end{table*}}
\def\btab{\begin{tabular}} 
\def\btabu{\begin{tabular}}             \def\etabu{\end{tabular}}
\def\btabx{\begin{tabular*}}            \def\etabx{\end{tabular*}}
\def\bbib{\begin{thebibliography}{999}} \def\ebib{\end{thebibliography}}
\def\bver{\begin{verbatim}}             \def\ever{\end{verbatim}}
\def\bca{\begin{cases}}                  \def\eca{\end{cases}}

\usepackage{comment}
\usepackage{ulem}
\usepackage{etoolbox}
\usepackage{algorithmic}
\usepackage[demo]{graphicx}
\usepackage{subcaption}
\usepackage[colorinlistoftodos]{todonotes}
\usepackage[colorlinks=true, allcolors=blue]{hyperref}
\usepackage{physics}
\newcommand{\E}{\mathbb{E}}
\renewcommand{\P}{\mathbb{P}}
\newcommand{\Q}{\mathbb{Q}}
\newcommand{\R}{\mathbb{R}}
\newcommand{\cF}{\mathcal{F}}
\newcommand{\indiq}{\mathbf{1}}
\newtheorem{theorem}{Theorem}
\newtheorem{proposition}[theorem]{Proposition}
\newtheorem{corollary}[theorem]{Corollary}
\newtheorem{lemma}[theorem]{Lemma}
\newtheorem{observation}[theorem]{Observation}
\newtheorem{example}[theorem]{Example}
\newtheorem{definition}[theorem]{Definition}
\newtheorem{property}[theorem]{Property}
\newtheorem{remark}[theorem]{Remark}
\newtheorem{claim}[theorem]{Claim}
\newtheorem{notation}[theorem]{Notation}
\newtheorem{fact}[theorem]{Fact}
\newtheorem{assumption}[theorem]{Assumption}
\providecommand{\keywords}[1]{\textbf{\textit{Keywords---}} #1}
\title{A survey of electricity spot and futures price models for risk management applications}

\author[1]{Thomas Deschatre}
\author[1]{Olivier F\'eron}
\author[1]{Pierre Gruet}

\affil[1]{EDF Lab Paris-Saclay and FiME, Laboratoire de Finance des March\'es de l'Energie, 91120 Palaiseau \footnote{thomas-t.deschatre@edf.fr, olivier-2.feron@edf.fr, pierre.gruet@edf.fr}}
\date{}

\begin{document}
	\maketitle
\begin{abstract}
This review presents the set of electricity price models proposed in the literature since the opening of power markets. We focus on price models applied to financial pricing and risk management. We classify these models according to their ability to represent the random behavior of prices and some of their characteristics. In particular, this classification helps users to choose among the most suitable models for their risk management problems.
\end{abstract}

\renewcommand\labelitemi{{\boldmath$\cdot$}}
\section{Introduction}
Almost 40 years have passed since the first electricity market was opened in Chile and 30 years since the landmark event of privatization and deregulation of the electricity generation industry in England and Wales. Since then, the opening up of markets has affected most developed countries; for example, through the total liberalization achieved in Western Europe in 2014. This deregulation has exposed energy players to market price fluctuations and, as with any financial risk, has encouraged them to implement management strategies that limit the effect of these risks on their financial statements. In this review, we are interested in models dedicated to financial valuation and risk management in electricity markets, that is, models that can represent both spot prices (seen as risks) and derivatives of which the most important are futures contracts (seen as the main hedging assets, since electricity is not storable). Moreover, we only focus on models of electricity prices, even if some models can be applied to other financial assets, such as other commodities.
 
\subsection{History and evolution of electricity price models in the literature}
\begin{figure}[h!]
\begin{tabular}{cc}
     \includegraphics[width=0.485\columnwidth]{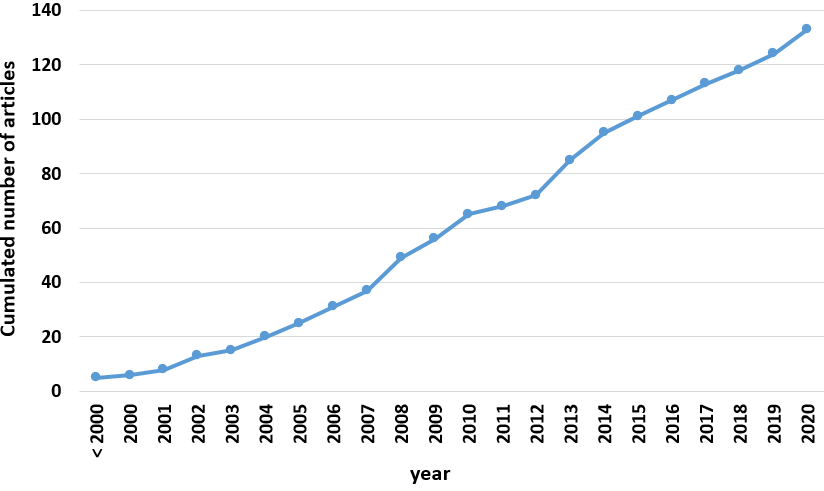} 
     &  
      \includegraphics[width=0.485\columnwidth]{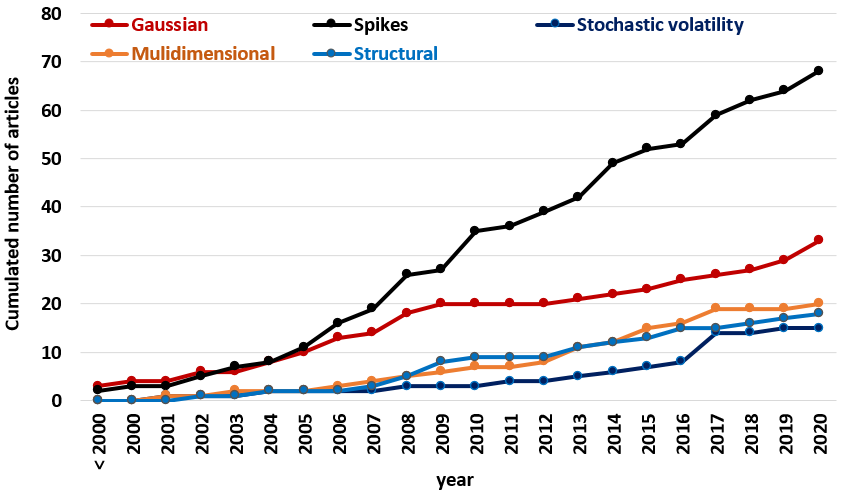}
     \\
     (a) &  (b) 
\end{tabular}
\caption{\label{fig:articles} Evolution by year of the cumulative number of papers in electricity price modeling for risk management: (a) global view and (b) view by model family.}
\end{figure}

\begin{figure}[h!]
\centering
\begin{tabular}{c}
\includegraphics[scale=0.6]{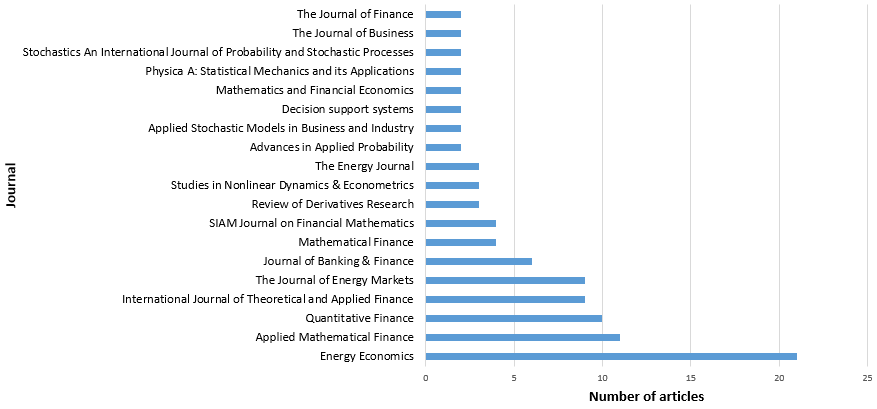} \\
\includegraphics[scale=0.6]{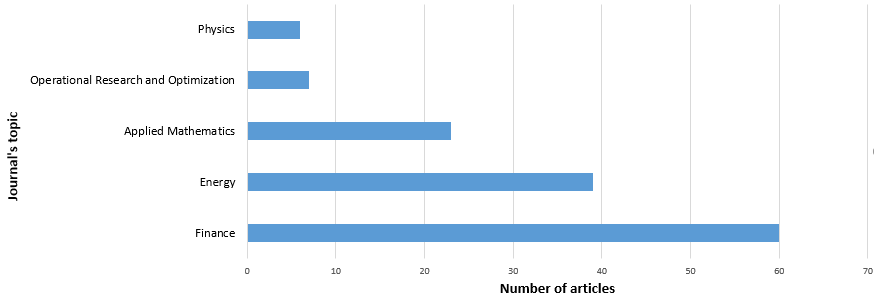} 
\end{tabular}
\caption{\label{fig:journal_articles} List of journals that have published more than two papers on electricity price modeling for risk management (top) and distribution of papers by journal topic (bottom).} 
\end{figure}

The first papers on the modeling of commodity prices for risk management applications were based on the convenience yield principle \cite{Gibson1990}, or they were based on the modeling of the interest rate curve \cite{Gabillon1991} that follows the principles of the Heath, Jarrow and Morton models~\cite{Heath1992}. It was not until the beginning of the 2000s that the first papers were published that really focused on electricity prices \cite{Bjerksund2000, Lucia2002}. Since then, this stream has proposed a great many models. Figure~\ref{fig:articles}.(a) shows relatively regular activity of an average of nearly 6 papers per year since 2000.
Several literature reviews are available that deal with electricity price models. For example, Carmona and Coulon~\cite{Carmona2014} is a review from 2013 on only the so-called ``structural models''. Weron~\cite{Weron2014} conducts a complete bibliographic review of models adapted to the short-term forecasting (1 day to a few weeks) of electricity spot prices.
Our objective here is to conduct a literature review of electricity price models that only focuses on financial valuation and risk management applications. The major difficulty lies in the great diversity of journals used by authors to propose electricity price models. We were able to gather about 170 papers (135 journal papers, 5 conference papers, 15 book chapters and 14 unpublished papers). The papers are published in 55 different journals on general topics such as energy, finance, applied mathematics, optimization and operations research, or physics (see figure~\ref{fig:journal_articles} (bottom)). Figure~\ref{fig:journal_articles} (top) shows the set of identified journals in which more than two papers on electricity price models appear. The journal \textit{Energy Economics} has the largest number of papers (21). Figure~\ref{fig:journal_articles} (bottom) shows that more than half of the papers are published in journals whose general theme is finance and financial mathematics. This trend illustrates the fact that the particular characteristics of electricity markets have prompted authors to propose new models that are different from those present in classical finance.
\subsection{Practicality of electricity price models}
There are multiple objectives that guide the proposal of an electricity price model in the context of financial valuation and risk management:
\bit
\item \textit{Practical use}: The models must be able to simulate prices and to price derivatives in a simple way within an acceptable timeframe. In particular, we seek to have quasi-explicit formulas for futures prices. This first point is emphasized by Carmona and Coulon~\cite{Carmona2014}: ``\textit{[...] commodities are mostly traded through forward contracts and the first challenge of a quantitative analysis is the computation of the term structure of forward prices.}'' ;
\item \textit{Good price representation}: The models must be able to reproduce the main characteristics of (spot and futures) prices ;
\item \textit{Price consistency}: The models must respect a coherence which is generally translated as the constraints induced by the absence of arbitrage opportunities ;
\item \textit{Identifiability}: Model parameters must be estimable from the available data. 
\eit
Most of the models presented in the literature are justified by the authors according to one or more of these elements. There are also several possible modeling choices that are partly related to the context of energy markets. For example, the question of whether to model the price arithmetically (i.e., the price itself) or geometrically (i.e., the logarithm of the price) arises as for any financial market price but even more so in the case of electricity, as specified by Barth and Benth~\cite{Barth2014}: \textit{``both additive and multiplicative models are of interest in energy forward markets''}. This is even more true nowadays when negative prices can be observed more and more frequently. Also, the available data, which are by nature on a time-discretized grid, lead to important modeling issues. Although the models, mostly in continuous time, can theoretically adapt to any time discretization, the specific problem of spot prices confronts the electricity market. Indeed, these prices appear at a daily granularity in the form of a set of hourly, half-hourly, or even quarter-hourly prices in some markets, for the following day. The reality of this market is therefore a large-scale process (of 24 hours, 48 semi-hours, or even 96 quarter hours)  observed on a daily time step. This large dimension can be extremely detrimental for valuation problems of complex derivatives. This is why most of the proposed models are one dimensional and often represent a daily spot price (seen as the average of hourly spot prices). However, in some papers \cite{Huisman2007, Hirsch2009}, the proposed process models an hourly spot price if the observed price process is of hourly granularity. 
\subsection{Proposed classification}
The classification of models most commonly observed in the majority of papers has two broad categories: models on spot prices and models on futures prices. This classification is historical and refers to the first two modeling approaches of Gibson and Schwartz \cite{Gibson1990} and Gabillon~\cite{Gabillon1991}, respectively. In more recent papers, we can find a third category that corresponds to the so-called ``structural models'', although these may fit into the category of spot price models. 
This classification does not appear to have a practical point of view for several reasons:
\bit
\item the targeted applications of valuation and risk management require a joint modeling of spot and futures prices, that is, a good representation of all the prices to which a manager is exposed;
\item from a spot (resp. futures) price model, one can find the corresponding dynamics on the futures (resp. spot) prices. In fact, a spot price model (resp. futures price) is equivalent to a futures price model (resp. spot price). 
\eit
 The classification proposed in this article is based on the capacity of the models to represent the random behavior of prices and some of their characteristics according to practitioner’s objectives. It is described below:
\bit
\item \textit{Gaussian factor models} in which the uncertainty is modeled by Brownian motions;
\item \textit{models to represent price spikes} which are by far the most numerous with a wide variety of approaches;
\item \textit{stochastic volatility models} which are a classical extension of Gaussian factor models;
\item \textit{multidimensional models} that seek to represent a set of prices in which one is the price of electricity;
\item \textit{structural models} which can naturally be seen as multidimensional models, but in which the link between variables is represented by complex functions linked to market fundamentals.
\eit
Gaussian factor models remain the preferred models of practitioners. They have many advantages, some of which are essential for practical use: they very often allow us to obtain closed formulas for the valuation of derivatives, and there are classic and efficient calibration methods. These characteristics explain, despite the simplicity of these models, the research’s constant use of this class of models, as illustrated in figure~\ref{fig:articles}(b). The representation of price spikes, which are present in the majority of electricity spot markets in the world, has guided most of the work on electricity price modeling. Indeed, figure~\ref{fig:articles}(b) shows that more than half of the published papers propose models that take price spikes into account. The questions raised, for this class of models, are very numerous. The explicit calculation of derivatives and the calibration of models are often an objective of the proposed models. But beyond that, the consideration of price spikes naturally means market incompleteness; and, as a result, some authors seek to study in depth the issues of probability change, risk premia, and hedging strategies. Stochastic volatility models are a natural extension of Gaussian factor models because they can depict thick tails of distribution as well as volatility smiles. This class of models is less developed for electricity prices and presents the same issues of market incompleteness.  Multidimensional models are used extensively in practice due to the need to value assets exposed to several markets. For example, the financial hedging of a power plant requires a model that represents both the price of electricity and the price of the fuels used to produce this electricity as well as possibly the price of CO$_2$ emissions. Also, many examples of spread options exist in the energy markets. In the case of electricity, they can represent a generation plant or an interconnection line between two countries. The most natural idea behind a multidimensional model is to duplicate the unidimensional models and consider a correlation on the noises. However, the observed link between the different prices modeled is much stronger than a simple correlation. Finally, structural models are a particularity of electricity price models, which is why we choose to make them a class by themselves. They are, by construction, multidimensional models but the link between electricity prices and the fundamental variables that build it is represented very finely to reproduce the way the electricity price is formed on the spot market. These strong and realistic links in particular can improve the management of spread options. However, the consideration of a structural model means market incompleteness, since the fundamental variables taken into account (electricity demand, available production capacities,...) are not hedgeable assets. 
\
\subsection{Organization of the article}
The rest of this article is organized as follows: The section \ref{sec:notations} introduces the notations as well as the fundamental relations between spot and futures prices. The sections \ref{sec:gaussiens} to \ref{section:modelesStructurels} constitute the bibliographical review according to the classification previously described. Finally, the section \ref{sec:conclusion} proposes a synthesis of this review and states our point of view on future work.

\section{Notation and preliminary remarks}
\label{sec:notations}
Before starting this review, it is necessary to introduce the notation that we use throughout the paper and to comment on the classic relations between spot and futures prices with various delivery periods.\par
In order to underline the time dependency of the quantities we study, the notation $X(t)$ refers to a deterministic function of time, while $X_t$ refers to the value of a stochastic process at time $t$. Moreover, we make a specific effort to keep the notation homogeneous throughout the paper. This is why the notation used to describe models can differ significantly from the original papers.
\subsection{Classic relations}
We denote by $f_t(T)$ the futures price that is quoted at time $t$ and delivered at time $T>t$. Like most of the research, we use the expression ``spot price'' for the price of a contract with immediate delivery and is denoted by $S_t$. Therefore, the first mathematical relation between spot and futures prices is:
\beq
\label{eq_relation_spot_forward}
S_t = \lim_{T \rightarrow t}f_t(T).
\eeq
Our definition of ``spot price'' is arguably applicable to commodities; definitions like ``the product with shortest maturity'' exist which, in this case, can vary from one commodity to another.
The second relation between spot and futures prices is given by the classic reasoning of the absence of arbitrage opportunity. We consider a filtered probability space $(\Omega, \mathcal{F}, (\mathcal{F}_t)_t, \mathbb{P})$, where $\mathcal{F}_t$ is the available information at time $t$. The classic absence of arbitrage opportunity argument leads to:
\beq
\label{eq_relation_forward_spot}
f_t(T) = \Ebb^{\Qbb}\left[ S_T | \Fc_t\right]
\eeq
where $\Qbb$ is the so-called ``valuation'' probability measure, which is risk-neutral, and means that discounted assets are $\Qbb$-martingales. Furthermore, the relation~\eqref{eq_relation_forward_spot} means that futures prices are $\Qbb$-martingales. Yet, in the case of commodities, the ``spot contracts'' are not tradable assets, and their price $S_t$ is not necessarily a martingale.
The models that are proposed in the literature start from either the spot prices or the futures prices. The two fundamental relations~\eqref{eq_relation_spot_forward} and \eqref{eq_relation_forward_spot}, which can also be recovered by using economic reasoning (see, e.g., A\"id et al.~\cite{Aid2018}), are commonly used to recover the whole set of electricity prices. 
\subsection{The specific case of electricity}
First, the spot prices of electricity (defined as the outputs of the spot market) are, in fact, futures prices with a maturity that spans between 12 and 36 hours, because the spot market only concerns deliveries during the hours of the next day. Yet,  the greatest part of the papers rely on the approximation~\eqref{eq_relation_spot_forward} or they consider a discretization of one day.
Second, the available futures products are binding contracts with a \emph{delivery period} that we call ``futures prices with delivery'' (FPD)\footnote{Depending on the markets, it is possible to trade forward or futures contracts. The difference is that for futures, there are regular margin calls between the contracting date and the delivery. The prices of forward and futures contracts are identical if the interest rate is deterministic. In this review, we are not making any distinction on the models based on the nature of contracts.}. We denote by $F_t(T_1,T_2)$ the price of the FPD delivering 1~MWh of electricity over the period $[T_1,T_2]$. The classic absence of arbitrage opportunity approach leads to a definition of a relation between the FPD and the forward curve:
\beq
\label{eq_relation_produit_forward}
F_t(T_1,T_2) =  \int_{T_1}^{T_2} w(s) f_t(s) ds,
\eeq
where $w(s) = \frac{1}{T_2 - T_1}$ if the contract is settled at $T_2$, and $w(s) = \frac{r e^{-rs}}{e^{-rT_1} - e^{-rT_2}}$ (where $r$ is the risk-free rate, assumed to be constant in this expression) if the contract is continuously settled during the delivery period. Here we stick to the above continuous summation formula presented by Benth and Koekebakker~\cite{Benth2008a}. There also is a discrete-time version of this relation, which is closer to practical considerations. \cite{Benth2008a} uses it in a general setting, and \cite{Feron2015, Feron2020a} use it in classic case. 
The essential link between spot prices $S_t$ (generally identified as bearing the risk) and the FPD $F_t(T_1, T_2)$ (identified as the hedging products) is the forward curve defined by the set of all unitary futures prices $f_t(T)$ for $T \ge t$. For this reason, all the models we gather in this review aim at either directly or not describing the forward curve, which is the cornerstone of the modeling of electricity prices.
Many research papers study the valuation principles of the derived products of forward curves, for instance Bj\"ork and Land\'en~\cite{Bjork2002} and Benth and Kr\"uhner~\cite{Benth2015a}. They define a general valuation framework and discuss the choice of a risk-neutral probability measure. However, electricity markets are incomplete, which many models acknowledge through price spikes, link between prices and exogenous variables that cannot be hedged, and so on. Thus, these models do not guarantee the unicity of the risk-neutral measure, and many research papers evoke the difficulty of defining this probability, while some other papers avoid discussing it. Moreover, the fact that electricity is not storable means that the consideration of the spot price as an underlying asset is possible, that is, one cannot buy, store, or sell it later. This can also alter the link between spot and futures prices as stated, for instance, by Vehvil\"ainen~\cite{Vehvilainen2002}: ``no analytical connection has been established between the [electricity] spot and forward prices'' or by Borovkova and Geman~\cite{Borovkova2006a} in 2006: ``electricity spot and forward prices are not closely related''.

\section{Gaussian factors models}
\label{sec:gaussiens}
This section presents all the models for which the price evolution comes from using Brownian motions and is important for several reasons:
\bit
\item It provides a history of electricity market modeling.
\item The models described in this section constitute the basic brick in the construction of the more advanced models seen in the following sections.
\item And, the class of these simple models is, still to this day, the preferred class of models for practitioners. Indeed, these models provide access to numerous closed formulas or approximations for the various derivative products that a risk manager must cover \cite{Clewlow2000, Aid2015}.
\eit
The first models of electricity prices were the Ornstein-\"Uhlenbeck and Black and Scholes models~\cite{Johnson1999}. However, when making a history of the models proposed for electricity markets, the first papers cited are the seminal work of Schwartz and his co-authors~\cite{Gibson1990, Schwartz1997, Schwartz2000} on commodities in general who introduced a key concept: the \textit{convenience yield}. This yield can represent, for example, storage costs or the preference to hold (and be able to consume) a commodity (see \cite{Carmona2004} for a summary description). In fact, these models follow the reasoning that came from the modeling of the interest rate curve, in particular the Va\v{s}\'i\v{c}ek~\cite{Vasicek1977} model. The set of models of type ``convenience yield'' is given in table~\ref{tab:modele_cy}.\\ 
Lucia and Schwartz~\cite{Lucia2002} were the first to apply this class of models specifically to electricity prices. They deal with the specificities of electricity due to its non-storable nature: the presence of a marked seasonality and the consideration of a delivery period in FPDs traded on the markets. They then formally propose a model that is equivalent to \cite{Schwartz1997, Schwartz2000} and applied it to seasonally adjusted spot prices. However, they leave aside the concept of the convenience yield and rely, as initiated in \cite{Schwartz2000}, on the idea of short-term shocks and long-term equilibrium.
\begin{table}[htb!]
\centering
\scriptsize
\begin{tabular}{p{15mm}p{40mm}p{42mm}p{40mm}}
\toprule
\scriptsize 
Model & 1 factor & 2 factors & 3 factors \\
\midrule
\centering Schwartz~\cite{Schwartz1997} & 
 $d \ln S_t = \alpha(\beta - \ln S_t)dt +\sigma dW_t$ 
&  $ \begin{aligned} 
\frac{dS_t}{S_t} & = & (\mu - \delta_t) dt + \sigma_1 dW^1_t \\
d \delta_t & = & \alpha (\beta - \delta_t) dt + \sigma_2 dW^2_t
\end{aligned}$ 
& 
$ \begin{aligned} 
\frac{dS_t}{S_t} & = & (r_t - \delta_t) dt + \sigma_1 dW^1_t \\
d \delta_t & = & \alpha (\beta - \delta_t)dt + \sigma_2 dW^2_t \\
d r_t & = & \kappa (\mu - r_t)dt + \sigma_3 dW^3_t 
\end{aligned}$ \\
\midrule
\centering Gibson and \newline Schwartz~\cite{Gibson1990}  & - & $\begin{aligned} d \ln S_t & = & (\mu - \delta_t) dt + \sigma_1 dW^1_t \\ d \delta_t & = & \alpha (\beta - \delta_t) dt + \sigma_2 dW^2_t \end{aligned}$ & -
\\
\midrule
\centering Schwartz and \newline Smith~\cite{Schwartz2000}
& - 
& $\begin{aligned} 
\ln S_t & = & X^1_t + X^2_t \\
dX^1_t & = & -\alpha_1 X^1_t dt + \sigma_1 dW^1_t \\
dX^2_t & = & \mu dt + \sigma_2 dW^2_t
\end{aligned}$
& - 
\\
\midrule
\centering
Lucia and Schwartz~\cite{Lucia2002}
&  
$\begin{aligned} 
P_t & = & S_t \mbox{ or } P_t = \ln S_t \\
P_t & = & \Lambda(t) + X_t \\
dX_t & = & -\alpha X_t dt + \sigma dW_t
\end{aligned}$
& $\begin{aligned} 
P_t & = & S_t \mbox{ or } P_t = \ln S_t \\
P_t & = & \Lambda(t) + X_t^1 + X^2_t \\
dX_t^1 & = & -\alpha X_t dt + \sigma^1 dW_t \\
dX_t^2 & = & \mu dt + \sigma^2 dW_t \\
\end{aligned}$ & - \\
\bottomrule
\end{tabular}
\caption{\label{tab:modele_cy} Models of ``convenience yield'' type. At date $t$, $r_t$ is the risk-free rate, $\delta_t$ is the convenience yield, and $\Lambda(t)$ is the seasonality. The $(W^n)_{n=1,\dots,3}$ are correlated Brownian motions.}
\end{table}
The model is written as:
\beq
\begin{split} 
P_t & = \Lambda(t) + X_t \\\
dX_t & = -\alpha X_t dt + \sigma dW_t
\end{split}
\label{eq:lucia}
\eeq
with $P_t = S_t$ (arithmetic case) or $P_t = \ln S_t$ (geometric case), $\Lambda$ as a deterministic function representing seasonality, $\alpha$ and $\sigma$ as two positive parameters. The authors then specify that the futures prices can be obtained analytically. In the geometric case, they first assume a particular change in probability that leads to a risk-neutral dynamic similar to \eqref{eq:lucia} by introducing a possible equilibrium value for the $X$ process: 
\[
dX_t  =  \alpha^*(\theta^* - X_t) dt + \sigma dW_t
\]
in which $\alpha^*$ and $\theta^*$ are the parameters of the risk-neutral dynamics. The futures prices are then obtained analytically using the Laplace transform:
\beq
f_t(T) = \exp \left\{\Lambda(T) + X_te^{-\alpha^*(T-t)} + \theta^*(1-e^{\alpha^*(T-t)}) + \frac{\sigma^2}{4\alpha^*} \left(1 - e^{-2 \alpha^* (T-t)} \right)\right\}.
\label{eq:lucia_forward}
\eeq
Thus, in this single-factor model, futures prices become almost constant for long maturities. Although this model provides an analytical formula for futures prices, it cannot be used in practice because the volatilities of long-maturity futures prices are almost zero. \cite{Aid2015} has already highlighted this problem. This general approach of the ``convenience yield'' type is also strongly questioned in empirical studies on energy commodities as, for example, in \cite{Carmona2004}, and is naturally questioned in the context of electricity which is not storable. The following papers~\cite{Hyndman2007, Schmeck2016, Latini2018} propose, as initiated in \cite{Schwartz2000}, the rewriting of the models in a more general way by talking about unobservable stochastic factors:
\begin{equation}
\begin{split}
\ln S_t & = \Lambda(t) + \sum_{n=1}^N X^n_t, \\
dX^n_t & = -\alpha_n X^n_t dt + \sigma_n dW^n_t.
\end{split}
\label{eq:modele_spot}
\end{equation}
 This formulation was already used in \cite{Karesen2000}. Further, there is equivalence between the Schwartz model~\cite{Schwartz1997} and the models \eqref{eq:modele_spot}. Indeed, Schwartz and Smith show in \cite{Schwartz2000} that their model is equivalent to that of Gibson and Schwartz~\cite{Gibson1990}. By making the same change of variable, Schwartz's 3-factor model can also be written in the form \eqref{eq:modele_spot}, with $N=3$.
The general form of futures prices is obtained, again, by using Laplace transform:
\[
f_t(T)  = \exp \left\{\Lambda(T) +\frac{1}{2}V(t,T) + \sum_{n=1}^N e^{-\alpha_n(T-t)}X^n_t \right\}
\]
with $$V(t,T) = \sum_{n=1}^N \sum_{n'=1}^N \rho_{nn'}\sigma_n \sigma_{n'} e^{-(\alpha_n + \alpha_{n'})(T-t)},$$
where $\rho_{nn'}$ is the correlation between the Brownian motions $W^n$ and $W^{n'}$. 
Concerning the calibration of these models, in \cite{Barlow2004} we find classical methods based on Kalman filtering. \par
At the same time, and still following the reasoning of interest rate modeling, other authors \cite{Gabillon1991, Clewlow1999a, Clewlow1999b} show interest in the approach of Heath, Jarrow and Morton~\cite{Heath1992} (HJM) which directly models a forward curve $(f_t(T))_{T\ge t}$. The general form is:
\beq
\label{eq:modele_hjmm}
\frac{df_t(T)}{f_t(T)} = \mu(t,T) dt + \sum_{n=1}^N \sigma_n(t,T) dW^n_t
\eeq
from where the solution can be obtained:
\[
f_t(T) = f_0(T) \exp \left\{ \int_0^t \right[\mu(s,T) - \frac{1}{2}\sum_{n=1}^N \sigma_n(s,T)^2 \left] ds + \sum_{n=1}^N \int_0^t \sigma_n(s,T) dW^n_s\right\}.
\]
Hinz et al.~\cite{Hinz2005} justify the use of HJM models in the context of electricity by reconciling interest rate models with electricity futures price models. Bjerksund et al.~\cite{Bjerksund2000, Bjerksund2010} were the first to apply the HJM models specifically to electricity prices\footnote{The working paper~\cite{Bjerksund2000} was available as early as 2000, but the work was only published in 2010 \cite{Bjerksund2010}.}.
They consider an HJM model with time-to-maturity dependent volatility functions, which is a form of \eqref{eq:modele_hjmm} with $N=3$ and
\begin{equation}
\sigma_1(t,T) = c, \quad \sigma_2(t,T) = \frac{a}{T-t+b}, \quad \sigma_3(t,T) = \left(\frac{2ac}{T-t+b} \right)^{\frac{1}{2}},
\end{equation} 
where $a$, $b$, and $c$ are positive constants. Here again, the authors face difficulties with electricity prices. In their framework, they study the dynamics induced by the FPD, i.e., with a delivery period. The model is studied in detail in the same article and then in \cite{Lindell2009}. Apart from these two papers, most of the works~\cite{Clewlow1999a, Clewlow1999b, Manoliu2002, Diko2006, Kiesel2009, Edoli2013, Kremer2020} that deal with the modeling of electricity prices with an HJM approach consider volatility functions of the exponential type:
\beq
\frac{df_t(T)}{f_t(T)} = \mu(t,T) dt + \sum_{n=1}^N \sigma_n e^{-\alpha_n (T-t)} dW^n_t,
\label{eq:modele_hjm}
\eeq
and in this framework, we can show that, in terms of the representation of uncertainties, HJM models are equivalent to the ``convenience yield'' type models given in the table~\ref{tab:modele_cy}\footnote{For perfect equivalence, we must consider, in the convenience yield models, time-dependent parameters in order to be able to stick to the initial forward curve.}. The main difference is in the representation of the initial curve where the HJM models are naturally adapted to fit the FPDs.
Models of type \eqref{eq:modele_hjm} are, to our knowledge, the preferred models of practitioners, and have been widely studied in the literature. The most important paper for this class of HJM models is undoubtedly that of Benth and Koekebakker~\cite{Benth2008a}\footnote{This paper is general and considers the dynamics driven by L\'evy processes for part of its results.}. Indeed, the authors give a complete theoretical analysis of this class of models in the specific context of electricity. In particular, they study in detail the dynamics induced on spot prices and FPDs. They show that the only way to keep a log-normal character for FPDs is to consider a volatility that is only a function of time, that is, which does not depend on maturity. The class of models that preserve log-normality is therefore very poor. In the case when papers consider more realistic volatility functions, integrating, for example, a maturity dependence to represent the Samuelson effect, and considering several Brownian factors, the dynamics induced on spot prices and FPDs are generally not log-normal, or even Markovian.
The first empirical studies appear in 2000 in \cite{Karesen2000} with a proposal for calibration by Kalman filtering that is necessary because of the non-Markovian character of both the spot price and FPD $F_t(T_1, T_2)$. However, the ``spot'' price that they consider is the weekly average. Studies have widely examined the problem of calibrating the model \eqref{eq:modele_hjm} that has $N$ factors \cite{Karesen2000, Manoliu2002}, or in more specific cases with three factors \cite{Diko2006}. The most studied model is the two-factor 
\begin{equation}
\frac{df_t(T)}{f_t(T)} = \mu(t,T) dt + \sigma_1 e^{-\alpha_1(T-t)} dW^1_t + \sigma_2 dW^2_t.
\label{eq:modele_2F}
\end{equation}
This model is equivalent to the models of Gibson and Schwartz~\cite{Gibson1990} and Schwartz and Smith~\cite{Schwartz2000}. It shows the importance of a constant volatility factor. Indeed, the absence of an exponential decrease in the second factor is essential to represent a non-negligible volatility for futures prices with long maturities.
The calibration of this model is studied in different ways: on marginal volatility and by maximum likelihood in \cite{Edoli2013, Feron2015}, by the development of an efficient statistical estimation procedure on FPDs in \cite{Feron2020a}, on implicit volatilities in \cite{Kiesel2009}, or by Markov chain Monte-Carlo method in \cite{Guerini2020}. Very recently, Fabbiani et al.~\cite{Fabbiani2021} calibrated the two-factor model on forward option prices in EEX markets by considering the model as the Schwartz and Smith~\cite{Schwartz2000} model. \cite{Koekebakker2005, Frestad2008} have discussed the number of factors to be used to represent electricity prices in the FPDs in the NordPool market. The authors observe that two to three factors explain about 70\% of the variations in futures prices. More recently, \cite{Feron2020b} estimates the optimal number of factors in both FPDs and spot prices for a model of type \eqref{eq:modele_hjm} for six European markets (Germany, Belgium, France, Italy, United Kingdom, and Switzerland). The optimal number of factors, estimated according to a Bayesian Information Criterion, is five for each market.
Kremer et al.~\cite{Kremer2020} are the first to look at the behavior of futures prices within a day. They conduct an empirical analysis of liquidity and volatility within a day and propose an estimation method that has a good representation of the relation between the realized variance within a day and the average duration between two price changes. 

\medskip
Some extensions exist to represent, for example, the seasonality of volatility, as observed by \cite{Benth2008a} in NordPool market data. \cite{Clewlow2000} introduces seasonality through a time-dependent function $t\mapsto\sigma(t)$ , but \cite{Audet2004} does so through a maturity dependent $T\mapsto\sigma(T)$.
Another natural extension is the nonlinear relations among Gaussian factors. For example, recent works by Kleisinger-Yu et al.~\cite{Kleisinger2020} and Ware~\cite{Ware2019} consider polynomial processes. The latter allow polynomial relations of Gaussian factors that retains the explicit formulas for futures prices. Further, the idea of structural models also consists, in general, in considering a nonlinear relation between Gaussian factors to enrich the representation of electricity prices, such as the Barlow model~\cite{Barlow2002}.

\medskip
In addition, there is some work based on other interest rate models. For example, Barth and Benth~\cite{Barth2014} use Musiela's approach, that is, they set $m=T-t$ and model the dynamics of $m\mapsto f_t(m)$ that represents the future price as a function of time to maturity. We can also cite a class of models based on the principles set out in \cite{Hinz2005} that allows an analogy to be made between futures products and zero-coupon bonds. In this context, several authors~\cite{Broszkiewicz2008, Biagini2015, Fanelli2016}\footnote{Note that Brotherszkiewicz-Suwaj and Jurlewicz~\cite{Broszkiewicz2008} consider a more advanced model with L\'evy processes to model jumps in the dynamics of futures contracts.} propose an HJM model on futures prices normalized by the spot price:
\beq \label{eq:normalizedspot}
P_t(T) = \frac{f_t(T)}{S_t},
\eeq 
so as to i) revert to a variable that is one at maturity, such as zero-coupon bonds, and ii) use the change of numeraire principles to value derivative contracts. However, the authors do not treat the case of zero or negative spot prices in this framework. 

\medskip
Finally, the last class of models~\cite{Borovkova2006a, Borovkova2006b, Broszkiewicz2006} is based on the Nelson and Siegel-type models~\cite{Nelson1987} that represent the forward curve by a mean level and a seasonality:
\beq
f_t(T) = \overline{f}_t e^{\Lambda(T)-\gamma_{t,T-t}(T-t)}
\label{eq:nelsonsiegel}
\eeq
in which $\overline{f}_t$ is the geometric mean of all futures prices (for any maturity), and $\gamma$ is a stochastic risk premium, which distorts the forward curve. 
Further, there is a class of models that is dedicated to the representation of FPDs as swaps. This idea was initiated in \cite{Benth2008b} and taken up again in \cite{Amadeu2020} where the authors consider a Gaussian one-factor model on each weekly FPD. However, these models do not allow us to obtain spot prices. Therefore, we consider them at the limit of the scope of this review.

\bigskip
In this first section, we have described the simplest models, both in the (Gaussian) representation of uncertainties and in their use to value and hedge derivatives. They are very close to the classical models used in finance and therefore make possible the use of the numerous results available for the classical financial markets and more particularly the interest rate markets. Further, this class of models generally presents two essential advantages for their practical use: the possibility of having derivative prices explicitly, and the existence of efficient methods of calibration on market data. However, these models are quickly limited in their ability to represent the specific characteristics of energy commodities and particularly electricity prices.

\section{Modeling of price spikes}
\label{sec:pic}
Johnson and Barz~\cite{Johnson1999} state as early as 1999 the need to represent positive and negative jumps as fundamental characteristics of the prices observed in the electricity markets of California, Australia, and Scandinavia. 
The existence of price spikes corresponds to a very high tension in the system, followed by a very rapid return to equilibrium, and are a consequence of the difficulty of storing electricity. There are both upward and downward price spikes, especially for regions with a high proportion of renewable generation. 
Their modeling is important in particular for the valuation of ``peak power plants'' which are called on when prices are very high (and modeled as call options with very high strikes), or for the valuation of flexibilities such as load shedding contracts or storage assets (which obtain maximum value during extreme price events).
In order to represent these price spikes, many papers adopt a reasoning similar to that of classical finance, namely, to generalize Gaussian models. This is classically done by replacing the Brownian motion(s) with more general L\'evy processes. These processes can represent jumps in a price series, as described, for example, in \cite{Tankov2003}. However, these models face a particular difficulty with electricity prices: the need to introduce a very strong mean reversion to represent a price spike. But this reversion is not representative of the mean reversion character of the electricity price in a normal, less tense, situations of the electricity system. Therefore, many studies propose several factors to dissociate the situations of tension (spikes) from normal situations. With the same objective, regime switching models have also been studied and are discussed in subsection~\ref{sec:regimeswitchingpic}. Finally, we will end this section by describing, in subsection~\ref{sec:clusterpic}, models that allow us to represent clusters of spikes, i.e. the fact that price spikes are not independent and appear most often in a grouped fashion.
\subsection{L\'evy processes}
\label{sec:processus_levy}
As previously stated, a natural way to represent price spikes is to replace Brownian motions with more general L\'evy processes. A first approach consists in generalizing the model \eqref{eq:modele_spot} to:
\beq
\begin{split}
    P_t & = \Lambda(t) + \sum_{n=1}^N X^n_t, \\
    dX^n_t & = -\alpha_n X^n_t dt + dL^n_t,
\end{split}
\label{eq:modele_spot_levy}
\eeq
where $P_t$ denotes the spot price (arithmetic case) or its logarithm (geometric case) on date $t$, and the $L^n$, $n=1,\dots, N$ are general L\'evy processes.
Bj{\"o}rk and Land{\'e}n~\cite{Bjork2002} have studied this model \eqref{eq:modele_spot_levy} in its general form to define the associated futures prices dynamics. In the arithmetic case when assuming the dynamics \eqref{eq:modele_spot_levy} are written under the risk-neutral probability, the futures prices dynamics are
\[
    df_t(T) = \sum_{n=1}^N e^{-\alpha_n(T-t)}\left(\sigma_n dW^n_t + \int_{\mathbb{R}} z \widetilde{\Pi}^n(dz, dt)\right)
\]
where $\sigma_n W^n$ is the continuous part of the process and $\widetilde{\Pi}^n$ the compensated Poisson measure related to the jumps in the process. The geometric case requires more technical conditions. If we assume the processes $L^n$ are independent and the dynamics \eqref{eq:modele_spot_levy} are written under the risk-neutral probability, we have:
\[
f_t(T) = f_0(T)\exp \left\{ \sum_{n=1}^N\int_0^t e^{-\alpha_n(T-s)}dL^n_s - \int_0^t \phi_n(e^{-\alpha_n(T-s)})ds \right\}
\]
with
\[f_0(T) = \exp \left\{\Lambda(T) + \sum_{n=1}^N \int_0^T \phi_n(e^{-\alpha_n(T-s)})ds \right\}
\]
and the dynamics of the futures prices are then given by
\[
df_t(T) = f_{t^-}(T)\left(\sum_{n=1}^N \sigma_n e^{-\alpha_n(T-t)}dW_t + \int_{\mathbb{R}} (\exp(z e^{-\alpha_n(T-t)})-1)\widetilde{\Pi}^n(dt, dz)\right)
\]
where $\phi_n$ is the characteristic function of the L\'evy process $L^n$. In the general case, the computation of these integrals is not explicit. Therefore, specific assumptions about the form of the L\'evy processes are needed, or numerical methods need to be used.
Benth et al.~\cite{Benth2007a} were the first to study a model of the type~\eqref{eq:modele_spot_levy} in its arithmetic form in the specific case of electricity. The authors used increasing independent increments (or subordinators) processes, potentially non-stationary, to have positive prices. They then obtained quasi-analytical formulas for the futures prices. \cite{Benth2012a} describes the calibration of this model.
Hess~\cite{Hess2020} proposes an extension that adds a stochastic average price level that accounts for the fact that some actors may have particular information (``insider information'') on this average level. A more detailed theoretical study of model \eqref{eq:modele_spot_levy} is done in \cite{Benth2008b} where the authors consider a part of the factors still driven by Brownian motions but the others are driven by general L\'evy processes. They study the possible changes of probability by using the non-storable character of electricity and then seek to explain the possible risk premiums (calculated by utility indifference) as a function of the behaviors of market participants. \cite{Benth2012e} also studies the risk premium in the model~\eqref{eq:modele_spot_levy} when it is in its geometric form (with both Brownian motions and L\'evy processes) by using an Esscher transform approach. In 2014, Benth and Schmeck~\cite{Benth2014b,Benth2014d} performed a similar study on the special case of a two-factor model in which the first was a mean-reverting process driven by a pure jump L\'evy process and the second was a Brownian motion. They then provided semi-analytical formulas for the values of call options on futures prices.
Also in 2014, Hess~\cite{Hess2014} studied the valuation of futures prices in the case of model \eqref{eq:modele_spot_levy} in its arithmetic form with the inclusion of future information. He extended the filtration and proposed in \cite{Hess2017b} optimal strategies for liquidating futures contracts in this model with a market impact.  
Benth et al.~\cite{Benth2013c} work on a slightly more general version of model~\eqref{eq:modele_spot_levy} with the form $S_t=g(t,X_t^1, \dots, X_t^N)$ in which $g$ is a continuous function, and the $X^n$ are driven by independent L\'evy processes. Their major contribution concerns the computation of the Greeks of options on the spot and futures prices.
Benth et al.~\cite{Benth2003} provide the most general version in which the non-homogeneous L\'evy processes are applied to the logarithm of the seasonally adjusted spot price along with a set of Brownian motions.
Crosby~\cite{Crosby2008} focuses on a similar model for futures prices and provides semi-analytical formulas for call option prices that makes the calibration of these option prices possible. This model is inspired by the work of Bj\"ork and Land\'en~\cite{Bjork2002} who propose a model on futures prices, jointly with a model of interest rates, of the HJM type, under real probability along with a Poisson measure $\Pi$:
\begin{equation} \label{eq:hjmpicsgeo}df_t(T) = f_t(T)\alpha(t,T)dt + f_t(T)\sigma(t,T)dW_t + f_{t^-}(T) \int_E \delta(t,z,T)\Pi(dt, dz).
\end{equation}
Further, Crosby~\cite{Crosby2008} considers an extension with nonlinear model on a factor of the form:
\[f_t(T) = H(t,Z_t, T)\]
with 
\[dZ_t = \alpha(t,Z_t)dt + Z_t\sigma(t,Z_t)dW_t + \int_E \delta(t,Z_{t^-}, z)\Pi(dt, dz).\] 
The function $H$ must then verify a certain partial differential equation for the futures prices to be martingale under the risk-neutral probability. The arithmetic version of \eqref{eq:hjmpicsgeo} is given by Benth et al.~\cite{Benth2019}: 
\[
df_t(T) = \left(c(t,T) - \lambda(t,T)f_t(T)\right)dt + \sigma(t,T)dW_t + \Phi(t,T)\int_{E} z \widetilde{\Pi}(dz,dt)
\]
where the authors establish the link with the dynamics of the FPD $(F_t(T_1, T_2))_t$. Similar to \eqref{eq:hjmpicsgeo}, which is a stochastic differential equation (SDE) on the futures prices, Hess' modeling~\cite{Hess2017a} is an SDE on the spot price. The model is arithmetic of the form
\[
dS_t = \left(\mu_t - \lambda S_t\right) dt + \sigma(S_t)dW_t + \int_{E}\delta(S_{t^-}, z)\widetilde{\Pi}(dt,dz)
\]
where $\mu_t$ is a stochastic equilibrium value. The author proposes constraints to ensure the positivity of spot prices as well as quasi-analytical formulas for futures prices.

\medskip
The calibration of models of type \eqref{eq:modele_spot_levy} is extensively studied in the literature. Most of the time, the calibration is done in several steps by treating the jump processes and the continuous part separately. For example, Deng and Jiang~\cite{Deng2005} perform a calibration of the model applied to increments of log prices. The estimation of the model is possible according to the marginal law of the increments from the stationary distribution. Meyer-Brandis and Tankov~\cite{Meyer2008} also do a complete estimation method for a two-factor model\footnote{The model differs slightly from \eqref{eq:modele_spot_levy} and is written $S_t = \Lambda(t)\left(X^1_t + X^2_t\right)$.} by using filtering methods, and Kl\"uppelberg et al.~\cite{Kluppelberg2010} do one for a three-factor model that uses EEX Phelix market data and is based on the extreme value theory. 
 
\medskip
As previously stated, the computation of futures prices from L\'evy processes is, in general, difficult and cannot be done with closed formulas. This constraint leads to difficulty, if not impossibility, of practical use. This is why the majority of the papers make law assumptions on the L\'evy processes in order to be able to perform these calculations. Therefore, papers have examined two classes of models: the class of compound Poisson processes and the class of Normal Inverse Gaussian (NIG) processes.
\subsubsection{Compound Poisson processes}
\label{sec:poisson}
The first papers consist of representing the jumps with Poisson processes. Johnson and Barz~\cite{Johnson1999} and Deng~\cite{Deng2000} propose considering a L\'evy process as the sum of a Brownian motion and a compound Poisson process. The proposed dynamics have one factor ($N=1$), and the model \eqref{eq:modele_spot_levy} is rewritten more simply as follows:
\beq
\begin{split}
\ln S_t & = \Lambda(t) + X_t, \\
 dX_t & = -\alpha X_t dt + \sigma dW_t + dM_t, 
\end{split}
\label{eq:1Fpoisson}
\eeq
where $M = (\sum_{i=1}^{\Pi_t} K_i)_t$ is a compound Poisson process that is related to the Poisson process $\Pi$ of intensity $\lambda > 0$ and to a series of random variables $K_i$ whose law is the same as the one of a random variable $K$. The intensity $\lambda$ represents the frequency of spikes per time unit, and $K$ represents the magnitude of these spikes. Moreover, \cite{Deng2000} studies a more general form in which the parameters of mean reversion, intensity, and volatility are time dependent with the latter also being stochastic. From a practical point of view, the time dependence of the parameters makes modeling the seasonality effects possible (e.g., the intensity $\lambda$ of the spikes indicates the spikes at classically tense moments of the power system like winter and peak hours).
Several papers~\cite{Benth2003,Cartea2005, Kjaer2008, Pawlowski2014} show the model \eqref{eq:1Fpoisson} in different forms that depend mainly on the jump law. For example, Kjaer~\cite{Kjaer2008} uses the compensated Poisson process $\tilde{M} = (M_t - \lambda \mathbb{E}(K)t)_t$: the choice is arbitrary and the model is of course equivalent to the model \eqref{eq:1Fpoisson}.

\medskip
In general, it is difficult to obtain an analytical formula for the futures prices from model \eqref{eq:1Fpoisson}, except for some particular laws on the magnitude $K$ of the jumps. For example, in the geometric case and with constant parameters, if we consider an exponential law for $K$ of density $x \mapsto \eta e^{-\eta x}$ with $\eta > 0$, then the futures prices can be obtained analytically:
\begin{equation} \label{eq:1FPoissonExp}
\begin{split}
f_t(T) = & \exp \Bigg \{ \Lambda(T) + X_te^{-\alpha^*(T-t)} + \theta^*(1-e^{\alpha^*(T-t)}) + \frac{\sigma^2}{4\alpha^*} \left(1 - e^{-2 \alpha^* (T-t)} \right) \\
 & \qquad + \frac{\lambda}{\alpha} \frac{\eta - e^{-\alpha(T-t)}}{\eta -1} \Bigg \}.
 \end{split}
\end{equation}
The exponential law is used in many papers~\cite{Deng2000, Villaplana2003, Kjaer2008, Benth2008b,Hambly2009, Kluppelberg2010, Pawlowski2014, Gonzalez2017}. Birge et al.~\cite{Birge2010} use a mixture of two exponential laws to represent positive or negative jumps. The normal law is also widely used \cite{Villaplana2003, Cartea2005, Seifert2007, Schmidt2008, Nomikos2008, Hambly2009, Nomikos2010a, Nomikos2010b, Bhar2013, Islyaev2015} but it does not give an analytical formula: there is still an integral to be calculated. \cite{Benth2003} uses the Dirac distribution, whereas \cite{Meyer2008} uses the Pareto law, and \cite{Kluppelberg2010} uses the the generalized Pareto law. We have considered in \eqref{eq:1FPoissonExp} the same change in probability as for the model of Lucia and Schwartz that leads to a risk-neutral probability described by the parameters $\alpha^*$ and $\theta^*$. We can observe that this formula is similar to \eqref{eq:lucia_forward} with the addition of an extra term that depends on the parameters of the Poisson process. Thus, as in \cite{Hambly2009, Nomikos2008, Nomikos2010b, Deschatre2018}, this additional term has a weak effect on the long-maturity futures prices. This is consistent with the fact that the spikes are short-term events. 

\bigskip
The estimation of the parameters of model \eqref{eq:1Fpoisson} on spot data can be done in several ways. In \cite{Cartea2005, Pawlowski2014} the authors first filter the jumps with an algorithm consisting of detecting returns with a value greater than three times the variance of the returns; they repeat this process until they only have returns below this threshold. The mean-reverting parameter $\alpha$ can then be estimated by linear regression, whereas the jump parameters are estimated on the filtered returns and the volatility on the unfiltered returns. Deng~\cite{Deng2000} uses a method of moments by matching the moments of the stationary distribution with the moments estimated on data. Kjaer~\cite{Kjaer2008} estimates the mean-reversion parameter using the empirical autocorrelation function and, once this parameter is fixed, the other parameters are obtained by maximum likelihood. An estimate of the risk premium is made in \cite{Cartea2005, Kjaer2008} to best fit the observed FPDs.

\bigskip
There are some extensions of the class of models \eqref{eq:1Fpoisson}. Geman and Roncoroni~\cite{Geman2006} propose an arithmetic model that can represent negative and positive jumps in which the sign of the jump depends on the level of the spot price. However, this model does not allow for the calculation of futures prices. Kegnenlezom et al.~\cite{Kegnenlezom2019} propose a modified version of the model \eqref{eq:1Fpoisson} that allows for an upper bound on prices. The calculation of futures prices is not explicit and requires approximations. Using a time change, Borovkova and Schmeck~\cite{Borovkova2017} introduce stochasticity to the mean-reversion, volatility, and jump intensity parameters. The time change may depend on an auxiliary variable. However, the latter paper does not study the valuation of futures prices. The time change principle is also used by Li et al.~\cite{Li2016} in the case of a CIR process with a compound Poisson component:
\begin{equation} \label{eq:CIR}
dX_t = \alpha(1 - X_t) dt + \sigma \sqrt{X_t}dW_t + dM_t.
\end{equation}
The time change is applied to the $X$ process using a subordinate L\'evy process. Further, Tong and Liu~\cite{Tong2017} use this principle to represent price spikes but without the use of the Poisson component in \eqref{eq:CIR}: the use of $(X_t^{\beta})_t$, $\beta \in \mathbb{R}$ instead of $(X_t)_t$ followed by a time change is sufficient. They obtain explicit formulas for futures prices and semi-explicit formulas for put option prices.
One of the main advantages of the one-factor model described above is, as for the Gaussian case, its Markovian property that is useful, if not essential, for both the estimation and the valuation of options, in particular swing options. However, we can state some limitations:
\bit
\item one of the main limitations, identified for example in \cite{Benth2012a}, is the presence of a single mean reversion parameter common to both the continuous and the spike parts. This leads to an underestimation of the mean reversion of the spikes which return to their initial level less quickly, and an overestimation for the continuous part. An overestimation of this mean reversion for the continuous part may also lead to an overestimation of the volatility to compensate for the strong mean reversion \cite{Benth2012a};
\item the model has the same drawbacks as in the Gaussian case: futures prices variations are totally determined by variations in the spot price, as already observed in \cite{Benth2003}; and, moreover, the volatility of long-maturity futures prices decreases too rapidly towards zero.
\eit
All these observations have led the authors to propose multi-factor models to represent all the characteristics of prices. As early as 2003, Villaplana~\cite{Villaplana2003} proposed a two-factor model of price dynamics:
\beq
\begin{split}
dX^1_t & = -\alpha_1 X^1_t dt + \sigma_1 dW^1_t, \\
dX^2_t & = -\alpha_2 X^2_t dt + dM_t. 
\end{split}
\label{eq:modele_villaplana}
\eeq
The case when $\alpha_1>0$ can represent both the mean-reverting characters in price spike situations and in normal situations. Further, the author also studies the case when $\alpha_1=0$ to obtain a significantly non-zero volatility for long-maturity futures prices. This model is then found in several papers on the case when $\alpha_1>0$ \cite{Nomikos2008, Benth2009a, Hambly2009, Nomikos2010a, Benth2013a, Benth2014c, Wei2020} or on the case when $\alpha_1 = 0$ \cite{Seifert2007, Islyaev2015}. Benth and Meyer-Brandis~\cite{Benth2009a} use techniques of enlargement of filtration to estimate the risk premium for $f_t(T)$ prices. Benth et al.~\cite{Benth2013a} use the same techniques for FPDs to confront market data. Benth and Ortiz-Latorre~\cite{Benth2014c} study a class of measure changes that allows for alterations in mean-reversion speeds in such a model. Birge et al.~\cite{Birge2010} take an original approach by applying a model similar to \eqref{eq:modele_villaplana} on peak demand and obtaining, via game theory, the spot price by calculating a Nash equilibrium. The proposed model adds a Brownian motion to the dynamics of $X^2$. Following the same idea, Branger et al.~\cite{Branger2010} propose adding a compound Poisson process to the Brownian motion in the dynamics of $X^1$ in order to consider both jumps and spikes on prices. \cite{Schmidt2008} uses a non-Markovian version of the process $X^2$ to represent the fact that some price spikes rise to their maximum value in several time steps: 
\beq
X^2_t = \sum_{i=1}^{N_t} K_i h(t-T_i, \gamma_i)
\eeq
where $(\gamma_i)_i$ is a series of random variables, $h(t,x) = e^{b(t-x)}$ for $0 \leq t < x$, and $h(t,x) = e^{-\beta(t-x)}$ for $t \geq x$ and $b, \, \beta \geq 0$. When $\gamma = 0$, this case is almost surely equivalent to the model \eqref{eq:modele_villaplana}.
Nomikos and Soldatos~\cite{Nomikos2010b} propose a three-factor model that solves the previously mentioned shortcomings. Indeed, the proposed model is geometric and is composed of the two factors described in \eqref{eq:modele_villaplana} (with $\alpha_1>0$) as well as a third factor driven by a Brownian motion:
\[dX^3_t  = \mu dt + \sigma_3 dW^3_t. 
\]
Thus, the model represents well the mean-reverting properties in spike and normal situations as well as significantly non-zero volatilities for long-maturity futures prices. 
The techniques for calculating futures prices in the case of these multifactor models are similar to the case of the one-factor model \eqref{eq:1Fpoisson}. For example, Hambly et al.~\cite{Hambly2009} propose numerical methods for the calculation of European and swing options in the case of the model~\eqref{eq:modele_villaplana}.

\medskip
Deschatre et al.~\cite{Deschatre2018} study the estimation of the parameters of the model~\eqref{eq:modele_villaplana} in a more general case where $X^1$ is an It\^o semi-martingale. The estimated values of the mean-reverting parameter $\alpha_2$ are very strong, which is consistent with the time series aspect. This strength means, as underlined for instance in \cite{Hambly2009}, a weak effect of the jump process on futures prices. The question of the number of factors arises in \cite{Gonzalez2017} where a Bayesian estimation method estimates a model whose first factor is Gaussian and the following factors are driven by compound Poisson processes with an exponential distribution for the jumps' magnitude. The number of factors is chosen according to the $p$-value of the Kolmogorov test performed on the Gaussian part reconstructed from the spot prices.

\medskip
As previously stated, there are very few models in this class that directly represent futures prices, since L\'evy's processes are mainly intended to represent price spikes in spot prices. However, a first attempt was made in 1999 when Joy~\cite{Joy1999} proposed an HJM-type model with the addition of a Poisson process to model jumps in electricity demand. The decay of the jump is not modeled by a mean reverting behavior but by a delay that creates a slightly different model:
\begin{equation} \label{eq:joydynamique}
df_t(T) = f_{t^-}(T)(\sigma(t,T)dW_t + dM_t).
\end{equation}
The Poisson component does not depend on the maturity, and when there is a jump on date $t$, it is cancelled in $t+\delta$, with the delay $\delta > 0$. However, this paper does not consider a risk-neutral probability. The valuations are made under the dynamics \eqref{eq:joydynamique}, while the futures prices are not martingale.
The generalization of the HJM model to the multifactorial case is, to our knowledge, done only by Crosby~\cite{Crosby2008}. The author proposes a more general model but studies an application of his model by considering Poisson processes.

\medskip
\subsubsection{Normal Inverse Gaussian processes}
\label{sec:nig}
The representation of L\'evy processes by Normal-Inverse-Gaussian (NIG) processes is commonly used in the financial literature for its simplicity.
 A L\'evy process $L$ is a NIG process if and only if $L_1$ has a NIG distribution whose characteristic function is:
\[
\phi(x ; \mu, a, \beta, \delta) = \exp\left(\mu x + \delta \left(\sqrt{a^2 - \beta^2} - \sqrt{a^2 - (\beta+x)^2}\right)  \right).
\] 
This law has several advantages:
\begin{itemize}
    \item its density and characteristic function are known, which facilitates the estimation and the pricing of options with the Fourier transform;
    \item the law is parameterized by only four reals (or even only three if we place ourselves in a martingale framework and constrain the trend term to zero). The parameter $\beta$ is the parameter that controls for the skewness while $a$ controls for the thickness of the distribution tails. 
\end{itemize}
Unlike the compound Poisson process, it can model jumps with infinite activity. We thus omit the diffusive part of the L\'evy process that can be represented by the small jumps.
To the best of our knowledge, in 2004 Benth and \v{S}altyt\.{e}-Benth~\cite{Benth2004}  were the first to test the relevance of NIG models for energy commodity prices, in particular for oil and gas; and in 2006 Collet et al.~\cite{Collet2006} applied these models to electricity prices. The proposed model is as follows: 
\beq
\label{eq:nigmodel}
\begin{split}
S_t & = \exp(\Lambda(t) + X_t) \\ 
dX_t & = -\alpha X_t dt + dL_t.
\end{split}
\eeq
It is then possible to have a semi-analytical formula for futures prices:
\[
\begin{split}
f_t(T) = \exp \Bigg \{ \Lambda(T) & + \delta \int_0^T \left(\sqrt{a^2 - \beta^2} - \sqrt{a^2 - (e^{-\alpha(T-s)} + \beta)^2}\right)ds\\
&+ \int_0^t e^{-\alpha(T-s)}dL_s 
 - \delta \int_0^t \left(\sqrt{a^2 - \beta^2} - \sqrt{a^2 - (e^{-\alpha(T-s)} + \beta)^2}\right)ds \Bigg \},
\end{split}
\]
with $|\beta + 1| < a$. 

\medskip
\cite{Benth2004, Collet2006} estimate the mean-reverting $\alpha$ parameter with a linear regression and the parameters of the NIG distribution with the residuals. \cite{Collet2006} shows that the NIG model can depict the spot prices of electricity well, and in particular the price spikes on the German EEX market. Further, \cite{Frestad2010} estimates the model directly on the FPDs observed on the NordPool market. The authors identify a risk premium and observe thick tails, which can be captured by the NIG process, but little skewness.
In 2014, Benth et al.~\cite{Benth2014a} generalize the model \eqref{eq:nigmodel} along two lines: (i) modeling the factor $X$ by a Continuous Autoregressive Moving Average (CARMA) process whose noise remains a NIG process and (ii) adding a second non-stationary factor to obtain non-zero volatilities for long-maturity futures prices. The use of CARMA processes, which are a continuous version of autoregressive moving average processes (ARMA), has several advantages. It enriches the dynamics of spot prices and does not complicate the calculation of futures prices compared to a classical Ornstein-\"Uhlenbeck process. The reader can refer to \cite{Brockwell2001} for more details on these processes. Benth et al.~\cite{Benth2014a} propose an original two-step estimation method: the estimation of the non-stationary factor by considering long-term FPDs (neglecting the effect of the short-term factor), and then the estimation of the CARMA factor by subtracting the estimated long-term factor from spot prices. More recently, Hinderks and Wagner~\cite{Hinderks2020b} calibrate a model similar to that of Frestad et al.~\cite{Frestad2010} on German spot prices. The model that they consider is a two-factor model: a first factor driven by a NIG process and a second driven by a Brownian motion or a compound Poisson process.
 
In 2013, Goutte et al.~\cite{Goutte2013} propose the generalization of the two-factor Gaussian model \eqref{eq:modele_2F}:
\beq
\frac{df_t(T)}{f_t(T)} = \mu(t,T) dt + \sigma_1 e^{-\alpha_1(T-t)} dL_t + \sigma_2 dW_t
\label{eq:modele_2F_nig}
\eeq
where $L$ is an NIG process. This model is taken up by the same authors in \cite{Goutte2014}. In these two papers, the authors propose a solution for hedging options by minimizing the $L^2$-distance between the payoff of the option and the hedging portfolio: this is called {\it Variance Optimal Hedging}. The choice of a risk criterion makes it possible to overcome the problem of market incompleteness induced by jumps in the price distribution. 
In 2014, Benth and Schmeck~\cite{Benth2014d} calibrate a similar model on EEX data, but in an arithmetic case and considering a second NIG process instead of a Brownian motion in \eqref{eq:modele_2F_nig}:
\[
df_t(T) = \sigma_1 e^{-\alpha_1(T-t)} dL^1_{t} + \sigma_2(T) dL^2_t.
\]
More recently, some papers have directly used the NIG processes to model the FPD $F_t(T_1, T_2)$. In 2018, Blanco et al.~\cite{Blanco2018} propose a model inspired by the work of \cite{Borovkova2006a, Borovkova2006b, Broszkiewicz2006} on Nelson-Siegel-type models~\cite{Nelson1987} that replace the Brownian motion with a L\'evy process in the equation \eqref{eq:nelsonsiegel}. In 2020, Piccirilli et al.~\cite{Piccirilli2020} propose a factorial model that is driven by NIG processes applied directly to FPDs and calibrate it on option prices that are computed by a Fourier transform. However, the two models mentioned cannot represent spot prices. 

\medskip
In 2012, Di Poto and Fanone~\cite{DiPoto2011} generalize the $N$ factors in a model where each factor follows a NIG distribution and is weighted by a time to maturity term of the form:
\[
\sigma_0 + \left(\sigma_1 + \sigma_2(T-t)\right)e^{-\alpha(T-t)}
\]
in which $\sigma_0$, $\sigma_1$, $\sigma_2$ and $\alpha$ are positive real numbers and $T \geq t \geq 0$. They are mainly interested in the estimation of the parameters and the number of factors that are done by using futures prices data in an independent component analysis.

\medskip
In an even more general framework, Barth and Benth~\cite{Barth2014} and Benth and Paraschiv~\cite{Benth2018} model futures prices using random fields that can be linkened to an infinite number of factors. The space dimension is time to maturity and is related to the Musiela parameterization.

\medskip 
NIG processes, as well as compound Poisson processes, are the most commonly used L\'evy processes for modeling electricity prices. However, it is of course possible to use other types of L\'evy processes: model \eqref{eq:modele_2F_nig} is also used with a Gamma Variance process by Goutte et al.~\cite{Goutte2014} and the class of $\alpha$-stable processes by Garcia et al.~\cite{Garcia2011} is used with a CARMA dynamics.
\subsection{Regime-switching models}
\label{sec:regimeswitchingpic}
The adoption of regime-switching models is also a natural and common way to represent extreme events. These models are very present in the literature and originate from the class of discrete time models, which makes it difficult to apply them to valuation and hedging issues. In particular, many papers exist with the sole purpose of modeling spot prices, such as \cite{Dejong2002, Huisman2003, Weron2004, Weron2005, Jablonska2011, Paraschiv2016a}. For example, Monfort and F\'eron~\cite{Monfort2012} describe a fully discrete-time model that obtains futures and options prices analytically. However, discrete-time models are outside the scope of this review because they do not easily identify hedging strategies.
 
Regime-switching models do indeed provide a good statistical representation of spot prices and in particular of spikes. Moreover, a comparison of several models conducted in 2007 by Bierbrauer et al.~\cite{Bierbrauer2007} on the EEX market confirms their interest. The principle of regime change is based on a state variable $U_t$ that defines the regime at time $t$. This state variable is generally driven by a Markov chain with transition probabilities from one state to another. In the case of electricity, the idea is to have a state corresponding to the spikes in prices. While in the case of L\'evy processes, the price spike is characterized by a very high mean reversion value, in the case of regime-switching models, the price spike event is characterized by a very high value of the probability of switching from the spike regime to a normal regime.  
As early as 2000, Deng~\cite{Deng2000} proposes a version of his model that integrated a regime shift to represent price spikes. The model is similar to model \eqref{eq:1Fpoisson} in which $M$ is the compensated Markov chain process that describes a state variable $U_t$ that has two values $U_t = 0$ and $U_t = 1$ at time $t$:
\[
\left\{
                \begin{array}{lcl}
                  \ln S_t & = &\Lambda(t) + X_t\\
                   dX_t & = &\alpha (\theta - X_t) dt + \sigma dW_t + \iota(U_{t^-}) dM_t\\
                    dM_t & = &-\lambda^{(U_t)}\delta(U_t)dt + dU_t \\
                   dU_t & = &\unb_{U_t = 0} \delta(U_t) d\Pi_t^{(0)} + \unb_{U_t = 1} \delta(U_t) d\Pi_t^{(1)}
                \end{array}
              \right.
\]
where $\iota(U_t)$ represents the random size of the jumps in regime $U_t$;  $\Pi_t^{(0)}$ and $\Pi_t^{(1)}$ are Poisson processes of intensity $\lambda^{(0)}$ and $\lambda^{(1)}$, respectively; function $\delta$ is defined by $\delta(0) = - \delta(1) = 1$; and $\unb$ is the characteristic function. A jump of the size of law $\iota(0)>0$ appears when we go from regime 0 to regime 1 and of size $\iota(1)<0$ when we go from regime 1 to regime 0. The laws chosen for the jumps in each regime are exponential. The change in regime a priori models two price levels (a high level $U_t = 1$ and a low level $U_t = 0$) more than spikes. However, if $\lambda^{(1)}$ is high, a positive jump is with very high probability followed by a negative jump that thus, models a positive price spike. Moreover, the calculation of futures prices requires the numerical solution of an ordinary differential equation.
Culot et al.~\cite{Culot2013} repeat the same principles of modeling the logarithm of the spot price as the sum of a diffusive part and a spike factor. The diffusive part is a model \eqref{eq:modele_spot} with three factors, and the spike factor is defined by four different states: 
\[
\ln S_t = \Lambda(t) + \sum_{n=1}^3 w_n X^n_t + \sum_{k=1}^3 \gamma_k {\bf 1}_{U_t = k}
\]
with $X^n$, $1 \leq n \leq 3$, Ornstein-\"Uhlenbeck processes weighted by a constant $w_n$. The state $U_t = 0$ is the state of no spike. The state $U_t = k$ corresponds to a spike of size $\gamma_k$ constant for $1 \le k \le 3$. The different states for $k \geq 1$ only distinguish the size of the spike. The spike shape is present if the probability of going from the $k$ state to the $0$ state is high for $1 \leq k \leq 3$. The process of spikes can be written as an affine process linked to a Poisson process to obtain a semi-analytical formula of the characteristic function of the process and thus of the futures prices. However, as for Deng~\cite{Deng2000}, it is necessary to solve a system of ordinary differential equations.

\medskip
To our knowledge, in 2005 Kholodnyi~\cite{Kholodnyi2005} was the first to propose a classical regime-switching model (i.e., with an unobservable state variable) that could obtain analytical formulas for futures prices. The diffusive part was still a \eqref{eq:modele_spot} model with one factor, and the spike factor was independent of the diffusive part:
\beq
\ln S_t = \Lambda(t) + X_t^1 + X_t^2
\eeq
where $X^1$ is an Ornstein-\"Uhlenbeck process, and $X^2$ is a piecewise constant process driven by a continuous-time Markov chain with two states: a state without a spike where the process is equal to zero, and a state with a spike where the process has a constant value of $\ln(\xi_{\tau})$ in which $\tau$ is the date of the change in state and $(\xi_t)_{t\geq0}$ is a collection of random variables greater than one in which the law depends on $t$.
When the distribution of $\xi_t$ does not depend on time, the futures prices are easily computed from the transition probabilities of the Markov chain and the average of the jumps. Let $(P_{i,j}(t,T))_{0 \leq i,j \leq 1}$ be the transition matrix of the Markov process between $t \geq 0$ and $T \geq t$ and $L(t) = \frac{dP(t,T)}{dT}|_{T=t}$ that is the infinitesimal generator. In the case where the law of $(\xi_t)_{t\geq0}$ does not depend on $t$, the futures prices are given by: 
\[f_t(T) = \lambda(t,T)f^{X^1}_t(T)\]
where $f^{X^1}_t(T) = \Ebb \left[ e^{ X_T^1} | \Fc_t \right]$ and 
\[
\lambda(t,T) = \left\{
    \begin{array}{ll}
        \lambda_t e^{\int_t^T L_{0,0}(u)du}  + \mathbb{E}(\xi_0) \int_t^T P_{1,0}(t,\tau)L_{0,1}(\tau)e^{\int_{\tau}^T L_{0,0}(u)du}d\tau  + P_{1,0}(t,T) & \mbox{si } \lambda_t > 1 \\
        \mathbb{E}(\xi_0) \int_t^T P_{1,1}(\tau,t)L_{1,0}(\tau) e^{\int_\tau^T L_{0,0}(u)du}d\tau + P_{1,1}(t,T) & \mbox{si } \lambda_t = 1 \\
    \end{array}
\right.
\]
In the case where the Markov process is homogeneous, that is if $P(t,T)$ is of the form $\tilde{P} \times (T-t)$, it is very simple to compute these integrals. The results are generalized in \cite{Kholodnyi2014} by considering a nonlinear function of $X_t^1$ and $X_t^2$. And in \cite{Kholodnyi2004, Kholodnyi2008}, he is particularly interested in numerical methods for the valuation of call options. The model explains the absence of spikes in the FPDs: the spike term is of the order of $1 + O(e^{-P_{1,0}(T-t)})$ (in the homogeneous case), that is, of the order of one if the time to maturity is large or if the probability of going from the jump state to the no jump state is large (which is the case for a spike). We then find effects similar to those observed in the case of L\'evy processes.

\medskip
J. Janczura is the author of several papers that deal with electricity price spikes, their estimation, and modeling in the framework of regime-switching models. In 2014, she proposed in \cite{Janczura2014b} a continuous-time regime-switching model in which regime-switching occurred only at discrete times. The model corresponds to the continuous version of the model proposed in 2010 in \cite{Janczura2010}. The price is written as the sum of a seasonality and a process described by three different states:
\[X_t = \left\{
    \begin{array}{ll}
        X^1_t & \mbox{si } U_{\lfloor t \rfloor} = 1 \\
        X^2_{\lfloor t \rfloor} & \mbox{si } U_{\lfloor t \rfloor} = 2 \\
        X^3_{\lfloor t \rfloor} & \mbox{si } U_{\lfloor t \rfloor} = 3 \\
    \end{array}
\right. ,
\]
with $\lfloor \cdot \rfloor$ the floor function, 
\[
    dX^1_{t} = -\alpha X_{t} dt + \sigma dW_t, \; t \geq 0,
\]
\[\log(X^2_{k} - c_2) \sim \mathcal{N}(\mu_2, \sigma_2^2),\; k \in \mathbb{N},\]
\[\log(-X^3_{k} + c_3) \sim \mathcal{N}(\mu_3, \sigma_3^2), \; k \in \mathbb{N},\]
and $U$ the discrete variable with three states directed by a discrete-time Markov chain.
The first state is the normal regime in which the dynamics of the seasonally adjusted price are an Ornstein-\"Uhlenbeck process. The second state depicts the positive jumps with a law of the sum of a Dirac with a positive value and a lognormal law. In the same way, the third state represents the negative jumps of law of the sum of a Dirac with a negative value and the opposite of a lognormal law. However, only a semi-analytical formula exists for the futures prices. The estimation is done using an EM algorithm. The estimation procedure is detailed in \cite{Janczura2014a} in the case of a slightly more general model.

\medskip
Another way to approach a regime-switching model is to consider the state variable observable, or dependent on an observable variable. This approach was proposed in 2002 by Davison et al.~\cite{Davison2002} who observed that price spikes occur frequently in the summer on the PJM market, while in the winter the spikes are less frequent and the price is more diffusive. They propose a two-regime model in which the probability of being in a given regime depends on the quotient between demand and the available capacity. The authors then calculate option prices from this model by assuming that traders are risk-neutral: the futures prices are given by the conditional expectation of the spot price under the historical probability.
Nomikos and Soldatos~\cite{Nomikos2008, Nomikos2010b} consider two or three-factor models in which the state variable is deterministic in \cite{Nomikos2010b} (having one specific value for winter and another for the rest of the year), or a function of the level of water stocks (a particularly important variable in the NordPool market) in \cite{Nomikos2008}. This assumption allows them to obtain analytical formulas for futures prices as well as for options prices. 
In 2016, Veraart~\cite{Veraart2016} proposed a CARMA(2,1)-type model driven by a two-regime process. The two regimes were modeled with L\'evy processes, and the state variable was determined by the level of the wind penetration index\footnote{The wind penetration index is defined as the ratio between wind generation and total electricity generation}. The model is then estimated for data from the German--Austrian area. The penetration index was also used in 2019 by Deschatre and Veraart~\cite{Deschatre2017} who proposed a model in which negative price spikes were represented by a Poisson process whose intensity was a two-state function that was determined by the penetration index.
\subsection{Modeling of spike clusters}
\label{sec:clusterpic}
In order to improve the representation of extreme events in spot prices, some papers try to model the fact that price spikes do not usually appear alone and are present in the form of clusters. This observation is made in many statistical studies, for example, \cite{Becker2013, Herrera2014}. These clusters cannot be represented by L\'evy processes with memoryless jump intensities and thus by the models referenced above, even when an inhomogeneous Poisson process is used. Proposals for models for electricity prices are very recent, although it should be noted that models based on time changes, for example, \cite{Li2016}, can handle spike clusters.
A solution proposed in \cite{Becker2013, Herrera2014, Callegaro2019} consists in using Hawkes processes~\cite{Hawkes1971}, that is, Poisson processes $\Pi$ whose intensity is stochastic, and of the form:
\begin{equation} \label{eq:hawkes}
    \lambda_t = \mu_t + \int_0^t g(t-s)d\Pi_s,
\end{equation}
which can also be written as
\[\lambda_t = \mu_t + \sum_{i=1}^{N_t} g(t - T_i)\]
in which $T_i$ are the instants of jumps. $\mu_t$ is the base intensity that is exogenous to the process and can for example represent a seasonality. The function $g$ is the kernel that quantifies the self-excitation of the process: when a jump occurs at date $t$, $dN_t = 1$ and the intensity increases by $g(0)$; at $t^{'} > t$, the effect of this kernel is $g(t^{'} - t)$. The most commonly used kernel is:
\begin{equation} \label{eq:kernelhawkesexp}g(t) = \alpha e^{-\beta t},
\end{equation}
$\alpha,\; \beta > 0$, $\alpha < \beta$, allowing to have $(\Pi_t, \lambda_t)$ Markovian when $t \mapsto \mu_t$ is a constant function. 
To the best of our knowledge, Callegaro et al.~\cite{Callegaro2019} are the only ones who study the modeling of futures prices with these Hawkes processes, and the parameters are estimated on these prices. The futures prices are modeled as a sum of factors depending on $t$ and $T$ whose continuous part is represented by a stochastic differential equation close to the CIR processes and the jump term by a marked Hawkes process (i.e., with a random jump size). The stochastic intensity which is of the form \eqref{eq:hawkes}--\eqref{eq:kernelhawkesexp} also account for the marking of the process: the intensity does not only increase by $\alpha$ when there is a jump but by $\alpha$ multiplied by the jump size. However, this term does not depend on the maturity and therefore a jump is simultaneous for all maturities. The estimation of the parameters of a Hawkes model can easily be done by means of a likelihood maximization (once the jumps are detected).

\medskip
An alternative present in \cite{Callegaro2019} for futures prices but also in \cite{Jiao2019} for spot prices is the use of a continuous branching process with immigration (CBI). These models also relate to the idea of self-excitation and are linked to a modeling by random fields. For example, in \cite{Jiao2019}, the spot price is defined as: 
\[
	S_t = \Lambda(t) + \sum_{n=1}^N X_t^n,
\]
where the $X^n$ follows the equation given below: for all $n \in {1,\ldots,N}$,
\[
	X_t^n=X_0^n+\int_0^t a_n(b^n-Y_s^n) ds +\sigma_n \int_0^t \int_0^{X_s^n} W^n(ds,du) +\gamma_n \int_0^t \int_0^{X_{s_-}^n} \int_{\R_+}\xi \widetilde{\Pi}^n(ds,du,d\xi),
\]
where $W^n$ are white noises on $\R_+^2$, and $\widetilde{\Pi}^n$ are Poisson measures compensated on $\R_+^3$. A major finding of Callegaro et al.~\cite{Callegaro2019} is that the CBI dynamics are rejected by a Kolmogorov test, unlike the dynamics of a Hawkes process.

\bigskip

The modeling of spikes, which are present in the spot price, can be important for the valuation of certain options and for modeling extreme risks. The use of L\'evy processes for this modeling, such as the compound Poisson process or the NIG process, is natural and allows for easy generalization of Gaussian models, and in particular factorial models. The estimation and valuation methods are then adapted to this new framework. An additional complexity appears for the calculation of futures prices in which formulas often require the calculation of integrals. Regime-switching models are also widely used to represent these price spikes (but not only). However, futures prices are often difficult to access which limits their use for risk management. Finally, a set of papers try to represent a stylized fact about spot price data: the presence of price clusters with non-uniformly distributed prices. These papers need to introduce an additional complexity that is quite costly for the calculation of futures prices. Finally, the more recent polynomial model of Ware~\cite{Ware2019} discussed in section~\ref{sec:gaussiens} is promising because it can represent price spikes while having simple calculations for a lot of derivatives as described by Benth~\cite{Benth2021}.

\section{Stochastic volatility models}
\label{sec:stovol}
Taking a stochastic volatility into account has two main advantages. First, it introduces asymmetry to the price distribution as well as fat tails. Thus, it better represents prices. The second advantage is the possibility to better characterize option prices in the market, such as the volatility smile. The proposed models for electricity prices are very similar to the classical stochastic volatility models. The first approach consists in extending the Schwartz model \eqref{eq:modele_spot} by considering a stochastic dynamic for the volatility parameter. At the end of this section, we present models based on semi-stationary L\'evy processes which, by their nature, allow us to represent stochastic volatility.
\subsection{Generalization of the Schwartz model}
The first stochastic volatility models date from 2000 \cite{Deng2000, Kellerhals2001, LariLavassani2001} and follow the principles of the classical Heston model:
\beq
\label{eq:modele_heston}
\begin{split}
dS_t & = \mu S_t dt + \sqrt{v_t} S_t dW_t,\\
dv_t & = \alpha_{v}(\theta_{v} - v_t)dt + \sigma_{v} \sqrt{v_t}  dW^{v}_t,
\end{split}
\eeq
in which $W^{v}$ is a Brownian motion correlated with $W$. The volatility is a CIR process that is in a Markovian framework with a volatility remaining positive. However, the model of Deng~\cite{Deng2000} is more general because it considers time-dependent parameters and price jumps that are common to the volatility process that incorporates them. However, futures prices in this general case are obtained with difficulty by solving ordinary differential equations. In the case of the model \eqref{eq:modele_heston}, in 2001 Kellerhals~\cite{Kellerhals2001} used the known results of the Heston model: after considering the classical change of probability using the Girsanov formula, the futures prices are calculated using the properties of affine processes. They are of the form:
\[
f_t(T) = S_t e^{A(t,T) v_t + B(t,T)}
\]
where the functions $A$ and $B$ are explicitly given in \cite{Kellerhals2001}.
Again in 2001, Lari-Lavassani et al.~\cite{LariLavassani2001} propose a similar model in which the square of the volatility has the following dynamics:
\[
dv_t = \alpha_{v}(\theta_{v} - v_t)dt + \sigma_{v} v_t^{\gamma}  dW^{v}_t
\]
in which $\gamma$ equals zero or one. This dynamic is incorporated in a model with one or two factors. The case $\gamma = 1$ corresponds to a GARCH(1,1) process. When $\gamma = 0$, the process can have negative values, but this problem is not treated by the author. The properties of affine processes can again be used to analytically obtain the futures prices. 
For each of the models proposed in \cite{Kellerhals2001, Deng2000, LariLavassani2001}, parameters can be estimated with Kalman filtering.

\medskip
In 2008, Hikspoors and Jaimungal~\cite{Hikspoors2008} proposed a general form for the volatility that they defined as a positive function of the value of a factor $Z$ following an Ornstein-\"Uhlenbeck dynamic whose Brownian motion was correlated to the one of the price. Since the model is general, the futures prices cannot be calculated exactly as only a partial differential equation is available. However, it is possible to have futures prices in an approximate way using an asymptotic development by assuming that the mean-reverting parameter of the $Z$ factor process is strong. 
Islyaev and Date~\cite{Islyaev2015} also propose a relatively general form for volatility of equation:
\[d\sqrt{v_t} = f_1(t,\sqrt{v_t})dt + f_2(t,\sqrt{v_t})dW^{v}_t.\]
The functions $f_1$ and $f_2$ are chosen so that the stationary distribution of volatility is a lognormal distribution (however, $f_1$ and $f_2$ are not specified). The authors emphasize that stochastic volatility models are easier to estimate than models with jumps and that they are simpler to use for pricing and hedging. Futures prices are given as an infinite sum (depending on the different moments of the process). An estimation by particle filtering is done on data of futures prices. 

\medskip
The principles of time change can also feature stochastic volatility. Li and Linetsky~\cite{Li2014} use a one-factor Gaussian Ornstein-\"Uhlenbeck model to consider a time change in the form $T_t = \int_0^t (a_u + Z_u) du$ where $Z$ is a CIR process. This time change allows us to introduce stochastic volatility into the model (as well as inhomogeneity in time). The futures prices are given in the form of an infinite series, and the model is directly calibrated to option prices. Note that the papers \cite{Li2016, Borovkova2017, Tong2017} described in the section~\ref{sec:poisson} also use a time change to represent a stochastic volatility.

\bigskip
More recent papers by Benth and his co-authors~\cite{Benth2013d, Benth2011a, Benth2013b} revisit the Barndorff-Nielsen and Shephard stochastic volatility model~\cite{Barndorff2001}. In \cite{Benth2011a}, Benth proposes adding stochastic volatility to the one-factor Schwartz model:
\beq
\begin{split}
S_t & = \exp(\Lambda(t) + X_t), \\
dX_t & = (\mu - \alpha X_t)dt + \sqrt{v_t} dW_t.
\end{split}
\eeq
The square of the volatility is represented by a multifactor model:
\[v_t = \sum_{i=1}^I w_i Y^i_t\]
with 
\[dY^i_t = -\kappa_i Y^i_t dt + dL^i_t,\]
$\kappa_i > 0$, and where the $L^i$ are subordinating L\'evy processes, that is, increasing. A semi-analytical formula is given for futures prices that involve the cumulative functions of L\'evy processes. The dynamics of the futures prices is given under certain integrability conditions as:
\[
df_t(T) = f_{t^-}(T) \left(\sqrt{v_t} e^{-\alpha(T-t)}dW_t + \sum_{i=1}^I \int_0^{+\infty} \left(e^{w_i \gamma(T-t, 2\alpha, \kappa_i) \frac{z}{2}}-1\right)\widetilde{\Pi}^i(dt,dz) \right)
\]
where $\widetilde{\Pi}^i(dt,dz)$ is the compensated Poisson measure associated with the process $L^i$ and $\gamma(s, a, b) = \frac{1}{a-b}\left(ae^{-as} - be^{-bs}\right)$ for $a \neq b$, $s \geq 0$. The value of futures prices is then:
\[
\begin{split}
f_t(T) = f_0(T)\exp \Bigg \{ \int_0^t \sqrt{v_s} e^{-\alpha(T-s)}dW_s & +\frac{1}{2}\sum_{i=1}^I w_i \gamma(T-t, 2\alpha, \kappa_i) Y^i_t \\
& - \sum_{i=1}^I \int_0^t \phi_i(- i \frac{w_i}{2} \gamma(T-s, 2\alpha, \kappa_i))ds\Bigg \}
\end{split}
\]
where $\phi_i$ is the characteristic function of $L^i$.
The futures prices are Markovian in the spot price and in each volatility factor and have, moreover, the same stochastic volatility $(\sigma_t)_t$. An interesting feature of the model is that the volatility factors are involved in the dynamics of the futures prices. 
Estimation of the stochastic volatility parameters is done in \cite{Benth2011a} on the residuals obtained after estimation of the seasonality function $\Lambda$ and the mean reversion parameter $\alpha$. The autocorrelation function gives the number of factors. The choice of the L\'evy factor model is made using the empirical density of the residuals. The risk premium is estimated by minimizing the Euclidean distance between the theoretical and observed FPDs. Benth~\cite{Benth2013b} studies this volatility modeling in the framework of the two-factor model of Schwartz~\cite{Schwartz1997}. The general multidimensional and multifactor case is studied in Benth and Vos~\cite{Benth2013d}, which we describe in section~\ref{sec:multidimensionnels}.
\subsection{Semi-stationary L\'evy processes}
\label{sec:lss}
In 2013, Barndorff-Nielsen et al.~\cite{Barndorff2013, Barndorff2014, Barndorff2015} proposed an approach based on semi-stationary L\'evy (LSS) processes. These were described in detail in \cite{Barndorff2013}. The proposed model is as follows:
\beq
\label{eq:lss}
\begin{split}
    P_t & = \Lambda(t) + X_t, \\
    X_t &  = \int_{-\infty}^t g(t-s)\sigma_{s^{-}} dL_s,
\end{split}
\eeq
where $P_t$ is the spot price (arithmetic case) or its logarithm (geometric case), $L$ is a L\'evy process, $g$ is a positive deterministic function with $g(t) = 0$ for $t \leq 0$, and $\sigma$ is a c\`adl\`ag process. Further, a more general, non-stationary version is described in \cite{Barndorff2013} that corresponds to Volatility Modulated L\'evy-driven Volterra processes in which the kernel $g$ is of the form $(t,s)\mapsto G(t,s)$, but is not used in practice. The model \eqref{eq:lss} depicts stationary spot prices to model price spikes with the non-Gaussian character of the returns and also to include a stochastic volatility. The kernel $g$ also reproduces different forms of autocorrelation function. This rather general model includes various classical models without stochastic volatility such as the one-factor model or the CARMA processes. Under certain conditions on $g$ and $\sigma$, the process $X$ is a semi-martingale. To allow the calculation of futures prices, the change of measure from $\P$ to $\Q$ is done by using a generalized Esscher transform. In the arithmetic form, the futures prices equivalent to the model \eqref{eq:lss} are given as:
\[
	f_t(T)=\Lambda(T) +\int_{-\infty}^t g(T-s)\sigma_{s_-}dL_s+\E_\Q(L_1)\int_t^T g(T-s)\E_\Q(\sigma_s|\cF_t)ds.
\]
In geometric form, the futures prices are given as the very general form:
\[
f_t(T) = \exp(\Lambda(T))\exp(\int_{-\infty}^{t}g(T-s)\sigma_{s^{-}}dL_s)\E_{\Q}(\exp(\int_t^T \Phi_{\Q}(\sigma_s g(T-s))ds) | \mathcal{F}_s)
\]
where $\Phi_{\Q}$ is the logarithm of the characteristic function of the L\'evy process under $\Q$. In the general case, the formulas are not explicit. 
One obtains explicit formulas in some particular cases, such as the case of constant volatility, that correspond to the models already studied. The authors also show that in the particular case of Brownian motion, and if the square of the volatility is written as $v_t = \int_{0}^t i(t-s)dU_s$ with $U$ as a L\'evy process and $i$ as a deterministic kernel, the futures prices are affine in the spot price and $v_t$ if and only if $g$ and $i$ are exponential functions, that is, we are working with Ornstein-\"Uhlenbeck processes. 
An empirical study is performed in \cite{Barndorff2013} on Phelix peak load data and an LSS model with constant volatility is calibrated on the data: the kernel is determined with the autocorrelation function and the stationary law of the maximum likelihood process. The CARMA(2,1) kernel with a symmetric NIG process for the stationary law also models the prices well. 

\medskip
The arithmetic LSS model \eqref{eq:lss} is also considered in 2020 by Rowi\'nska et al.~\cite{Rowinska2020} with the addition of a long-term factor which is a L\'evy martingale process. The estimation of the model is then carried out on spot prices and FPDs in Epex and EEX. This step leads in particular to testing different distributions to represent the increments of the L\'evy processes, as in Veraart~\cite{Veraart2016}. The authors consider the case of stochastic volatility when the LSS process is a Brownian motion of the form $v_t=\int_{-\infty}^t e^{-\delta(t-u)}dV_u$ with $V$ as a L\'evy subordinator that is independent of $B$. The estimated parameter $\delta$ is statistically significant, and the authors conclude that its inclusion improves the quality of the model. The links between the LSS process and the wind penetration index and residual demand are then examined. 

\medskip
A generalization of the LSS model is made in \cite{Barndorff2014} (and in \cite{Barndorff2015} for the multidimensional version) where futures prices are directly modeled as random fields with a time dimension and a time to maturity dimension.\par
In 2017, Bennedsen et al.~\cite{Bennedsen2017} also used LSS models to measure the roughness\footnote{The roughness property of a process is related to a particular behavior of its variogram, it is defined in \cite{Bennedsen2017}.} of the trajectories of the spot price. In their model, spot prices are driven by the sum of a factor such as the one appearing in the equation~\eqref{eq:lss} and a Wiener integral. The estimation of one of the volatility parameters leads to the conclusion that prices in several markets are rough.

\bigskip
As for the models with price spikes, the models with stochastic volatility are a natural extension of the Gaussian models. The introduction of a stochastic volatility classically allows in finance the representation of fat tailed as well as volatility smiles for option prices. The first models studied in this section fit into this framework with natural extensions of the classical models used for electricity. Multifactor modeling is also taken up for the modeling of volatility dynamics. The very general class of semi-stationary L\'evy processes provides a very complete theoretical framework that encompasses most of the models studied so far in this review. However, this framework induces very general formulas for futures prices, and their computation is only possible in some particular cases encountered so far. In general, the introduction of stochastic volatility complicates the calculation of futures prices.

\section{Multidimensional models} \label{sec:multidimensionnels}
The previous sections describe models that attempted to represent the price of electricity in a single market. In practice, however, the practitioner very often is confronted with assets exposed to several markets (electricity prices in several interconnected markets, or electricity and fuel prices). In particular, the problem of the representation of spreads has grown in importance during the liberalization of energy markets. The risk management activities of electricity producers encourage them to represent the payoffs of their generating plants in order to hedge them. The gain of a power plant can be seen, under certain simplifying hypotheses (absence of dynamic constraints, absence of start/stop costs, etc.), as the payoff of an option on the spread between electricity and the commodities that enable it to be produced. On the other hand, the modeling of the electricity prices in two interconnected markets can show the value of an interconnection line.
Therefore, the very early literature naturally approached the joint representation of several prices in energy models; such as Deng~\cite{Deng2000} who, as early as 2000, proposed a multidimensional model that featured the price of a fuel along with the price of electricity; and Deng et al.~\cite{Deng2001} who, in 2001, studied the problem of the valuation of generating plants and interconnection lines. Since then, many papers have addressed multidimensional models. Of course, structural models are by nature entirely part of the class of multidimensional models. However, the size and complexity of these models make their application to pricing and hedging problems difficult. In this section, we describe the set of multidimensional models that follow the reasoning of the so-called ``reduced-form'' models. 
 \par
The most natural approach to represent multiple commodity prices is to duplicate the one-dimensional models and consider correlations between the noise driving the processes. This class of models is studied in section~\ref{sec:modele_multidim_correlation}. We then present, in the subsection~\ref{sec:modele_multidim_variable_commune}, an approach based on the introduction of a random variable coupling several prices. This variable can enter into each of their dynamics, or act as a hidden variable governing them. Another approach, widely developed in the statistical literature, is based on the principles of cointegration. This approach has also been proposed in the context of electricity, the work is presented in section~\ref{sec:modele_multidim_cointegration}. 
\subsection{Correlation models}
\label{sec:modele_multidim_correlation}
As stated above, Deng~\cite{Deng2000} first proposed a multidimensional model in which he considers the model \eqref{eq:1Fpoisson} for both the electricity price and the fuel price. Further, he considers correlated Brownian motions that direct these two processes. In 2001, he and his co-authors~\cite{Deng2001} proposed Gaussian models on futures prices:
\[
\frac{dF_t^d(T)}{F_t^d(T)}=\mu^d dt + \sigma^d dW_t^d, \quad d=1,2,
\]
where $F_t^d(T)$ denotes commodity futures prices $d$ in $t$, $W^1$ and $W^2$ are two Brownian motions correlated with $\rho\in(-1,1)$, $\mu^d$ is a real, and $\sigma^d$ is a positive real. The authors then use Margrabe's classical formula~\cite{Margrabe1978} to value geographical spread and spark spread\footnote{The spark spreads are the spreads between electricity and gas prices.} options.
The same authors also propose an Ornstein-\"Uhlenbeck type model:
\[
dF_t^d(T)=\kappa^d\left(\mu^d(t)-\log(F_t^d(T))\right) F_t^d(T) dt + \sigma^d(t) F_t^d(T) dW_t^d, \quad d=1,2.
\]
The authors do not propose a method for estimating parameters, but they give a fairly complete set of formulas for valuing spark spread and geographic spread options. 
In the same idea, Meyer-Brandis and Morgan~\cite{Meyer2014} extend the one-dimensional model previously proposed by Meyer-Brandis and Tankov~\cite{Meyer2008} to the two-dimensional case by simple duplication:
\[
    S_t^d=\Lambda^d(t)(X_t^{d,1}+X_t^{d,2}), \quad d=1,2,
\]
with
\begin{align*}
        dX_t^{d,1}&=-\lambda_1^d(\mu_e-X_t^{d,1})dt+\sigma_d dW_t^d,\\
        dX_t^{d,2}&=-\lambda_2^d X_t^{d,2}dt + dL_t^d,
\end{align*}
in which $W^d$ is a Brownian motion and $L^d$ is a pure-jump process. They then consider a correlation between the Brownian motions, $W^1$ and $W^2$, and then study several examples of copula to link the L\'evy processes. They carry out a detailed estimation and simulation study on UK electricity and gas prices in order to value options on the spark spread.

\medskip
Gaussian models are also discussed in Carmona and Durrleman's literature review~\cite{Carmona2003} which presents spreads in the financial, commodities, and energy markets and discusses models for valuing them. They use the classical Ornstein-\"Uhlenbeck's models on the logarithms of spot prices, as well as the following HJM-type model:
\[
        \frac{dF_t^d(T)}{F_t^d(T)}=\mu_d(t,T)dt +\sum_{n=1}^N \sigma_{d,n}(t,T)dW_t^n, \quad d=1,2.
\]
Considering the same Brownian motions for the two processes $F^1(T)$ and $F^2(T)$ is another way to write two correlated models.
Still with the aim of valuing spread options, unlike the papers cited above Benth and \v{S}altyt\.{e}-Benth~\cite{Benth2006} propose directly modeling the spread 
rather than each of its two components. It is thus a univariate model of a multidimensional problem. In this case, Benth and \v{S}altyt\.{e}-Benth~\cite{Benth2006} propose representing $(Sp_t)_t$ with the model \eqref{eq:1Fpoisson} and obtain semi-explicit formulas for the valuation of spark spreads using Fourier analysis techniques.

\medskip
In 2007, Hikspoors and Jaimungal~\cite{Hikspoors2007} were interested in valuing spread options by duplicating a two-factor Gaussian model of type \eqref{eq:modele_spot} for each commodity. They considered the correlations between Brownian motions that drove the dynamics and then detailed the probability changes, the calculation of futures prices, and the valuation of spreads, which can be obtained explicitly.
Benth and Zdanowicz~\cite{Benth2016} also study the pricing of spread options based on a model duplicated for each commodity. In their case, the proposed model is a Volatility Modulated Volterra model. The authors then consider a bivariate L\'evy process that includes two correlated Gaussian components and a very general bivariate L\'evy measure. Semi-explicit formulas are then given for the pricing of futures prices and spread options. 

\medskip
While the papers presented above focus on the modeling of the spreads between two commodities, some papers aim at depicting a larger process to model more than two commodities simultaneously. For example, the three-price case of commodities allows the representation of a production plant that emits carbon dioxide. Its gains can then be represented as payoffs for three-variable spread options that integrate the price of CO$_2$ emissions.
Benth and Vos~\cite{Benth2013d} propose a model for the joint representation of $D$ commodities and focus on the presence of jumps, mean reversion, and stochastic volatility. Their $D$-dimensional price vector $S$ is given by:
\[
        S_t=\Lambda(t)\exp\left(\sum_{n=0}^N X_t^n\right),
\]
where the processes $X^n$ verify:
\[
\begin{split}
        dX^0_t & = AX^0_t dt + \Sigma_t^{1/2}dW_t. \\
        dX_t^n& = (\mu_n+B_n X_t^n) dt +\eta_n dL^n_t, \quad n = 1, \dots, N.
\end{split}
\]
They consider the square of the volatility $\Sigma_t$ of the process $X^0$ to be stochastic:
\[
        d\Sigma_t=\sum_{m=1}^M \omega_m Y_t^m,
\]
where
\[
        dY_t^m =(C_mY_t^m+Y_t^mC_m')dt + d\widetilde{L}^m_t.
\]
The $L^n$ and $\widetilde{L}^m$ processes are L\'evy subordinators. The authors further consider that $N<M$ and that the processes $L^n$ and $\widetilde{L}^n$, for $n\leq N$, have simultaneous jumps. They use this model after a change of probability to value options on spot and futures prices. 
In the same spirit, Grine and Diko~\cite{Grine2010} also propose a $D$-dimensional model whose component $d$ is:
\[
    S_t^d=\left(\sum_{d'\in\mathcal{D}_i} A(d, d') S_t^{d'}+\eta_d(t)\right)\varpi_d(t)\exp(b_d(t)X^d_t),
\]
where $X^d$ is a L\'evy diffusion that incorporates drift, volatility, and stochastic jump intensity. In the equation above, the term $\eta_d$ represents annual seasonal patterns and the term $\varpi_d$ represents weekly patterns; and for each $d$, $\mathcal{D}_d$ is a subset of $\{d'=1, \dots D: d' \neq d \}$. The terms $A(d,d')$ represent a linear relationship between commodity prices. Under a risk-neutral probability, the authors obtain a similar equation for futures prices and use it to value spreads through semi-explicit formulas. Contrary to Benth and Vos~\cite{Benth2013d}, Grine and Diko~\cite{Grine2010} briefly propose a calibration procedure, as well as a complete numerical study of the German and French markets.

\medskip
Conversely, in 2013 Edoli et al.~\cite{Edoli2013} proposed simple models to describe several commodities by focusing on the estimation procedure. The model does not focus on spot prices, but on homogeneous FPDs (i.e., contracts with the same delivery period $[T_1~;~T_2]$). Each of the $D$ FPDs follows the dynamics of the model by Kiesel et al.~\cite{Kiesel2009}:
\[
        \frac{dF_t^d(T_1,T_2)}{F_t^d(T_1,T_2)}=e^{-\lambda_d(T_1-t)}\sigma_d^1 dW_t^{d,1}+\sigma_d^2dW_t^{d,2}, \quad d=1, \dots, D.
\]
The authors then discuss several calibration methods of which one is a two-step estimation by market and is followed by an estimation of the correlations between markets.  

\medskip
More recently, Barndorff-Nielsen et al.~\cite{Barndorff2015} have approached multidimensional modeling with a very general framework that uses Ambit fields applied to futures prices. In this framework, the authors propose an expression for spot prices and obtain, under certain technical conditions, semi-explicit formulas for spread option values. They also address in depth the question of the change of probability. Further, Di Persio and Perin~\cite{Dipersio2015} address in depth the numerical aspects by comparing Ambit models with German EEX prices.
\subsection{Coupling factor models} 
\label{sec:modele_multidim_variable_commune}
Several papers use a random variable by coupling several processes in order to introduce a link between them; each process can also have its own dynamics, but this coupling variable makes it possible to display behaviors common to the processes. It can either enter into price dynamics, or constitute a hidden variable that governs price evolutions.\par 

\medskip
An example of a model in this class is that of Frikha and Lemaire~\cite{Frikha2013}: after performing statistical analyses on series of spot prices of electricity (Powernext) and gas (NBP), the authors propose the following model:
\begin{align*}
        S_t^1 & =\Lambda^1 (t) + X_t^1+Z_t, \\
        S_t^2 &=\Lambda^2(t) + X_t^2+Z_t,
\end{align*}
with
\[
dX_t^d=-\lambda_d (X_t^d-\mu_d)dt + \sigma_d(X_t^d,\theta_d) dW_t^d, \quad d = 1,2.
\]
The $Z$ component is common to both prices and has the following dynamics:
\[
        dZ_t=-\lambda_Z Z_t dt + \sigma_Z dW_t^Z.
\]
The Brownian motions are independent of each other. The function $\sigma_d$ is chosen so that the process $X^d$ is an ergodic diffusion of a given invariant probability and is parameterized by $\theta_d$. For example, the fat tailed distributions can be chosen to represent jumps.\par 

\medskip
In 2020, Christensen and Benth~\cite{Christensen2020} were interested in representing electricity prices in two interconnected countries. The authors sought to account for interconnectedness, that is, the non-zero probability event when prices are identical. A correlation is not sufficient to represent the fact that prices can be strictly equal over several time periods. They then propose a model based on two regimes: a regime with saturated interconnection in which prices are different and driven by correlated processes; a regime where prices are identical with a value given by a weighted average of the values of these two processes. The regime change is driven by the sign of a hidden variable modeled by an Ornstein-\"Uhlenbeck process with seasonality. The authors then give explicit formulas for futures prices and transmission rights.
In addition, they provide a procedure to estimate the model using filtering methods and then use the model to price transmission rights on the interconnection between France and Germany.
\par 

\medskip
Also in 2020, Gardini et al.~\cite{Gardini2020a} constructed prices driven by Brownian motions to which time changes were applied. These time changes were L\'evy processes that depend on each other that thus allowed the introduction of a link between the modeled prices and the induction of a delay phenomenon. The authors explain that this modeling could be used to represent prices in different countries and the possibility that an event in one of the two countries could propagate to the other with a time lag. They give several concrete examples of such models and present an application to the pricing of an option on the difference between French and German futures prices as well as to the pricing of spark spreads. Gardini et al.~\cite{Gardini2020b}, later in 2020, extended this paper by considering NIG processes in the same time-varying framework; the targeted applications were the same, and an efficient simulation algorithm was proposed for models using NIG processes.

\subsection{Cointegration models}
\label{sec:modele_multidim_cointegration}
The cointegration principle is based on the stationary representation of a linear combination of two processes, although the individual depiction of each process is non-stationary. It is quite natural to use this approach in the case of energy markets. For example, the observed data for the futures prices of some interconnected markets show that their difference oscillates around a certain long-term value. Such behavior cannot be reproduced with two Brownian diffusions, for example, unless they are perfectly correlated. However, in the case of electricity that is not storable, the spot price does not need to respect the martingale constraint under the risk-neutral probability. Therefore, accounting for stationary models as with cointegration to represent spot prices is possible and so is using it for the valuation of derivatives. 
In some sense, the article of Frikha and Lemaire~\cite{Frikha2013} of section \ref{sec:modele_multidim_variable_commune} could have been inserted in this section as it features a stationary difference between two spot prices. Yet, each of those two spot prices are already stationary, which is why the model does not meet the definition of cointegration.
\par 

\medskip
One of the first attempts at modeling with cointegration dates from 2011 and was proposed by Jaimungal and Surkov~\cite{Jaimungal2011}. They sought to represent the seasonality, jumps, cointegration, and mean reversion observed in commodity prices. They proposed a multidimensional $N$ factor model for $S$ in which the vector of $D$ commodity spot prices was jointly represented in:
\[
        S_t=\exp\left(\Lambda(t)+BX_t\right),
\]
with
\[
        dX_t=-K X_{t_-}dt + dL_t,
\]
where $\Lambda(t)$ is a deterministic vector, $B$ is a $\R^{D\times N}$ matrix, $K$ is a $\R^{N\times N}$ matrix, and $L$ is a $N$-dimensional L\'evy process. Cointegration could be reproduced by this model since the risk $L$ is common to all components of $S$. The authors also expressed futures prices with this spot price model and then priced European, American, and barrier options. The formulas were semi-explicit but the numerical methods were applied to illustrate the performance of the model for pricing.\par 

\medskip
In 2012, Nakajima and Ohashi~\cite{Nakajima2012} generalized the Gibson and Schwartz~\cite{Gibson1990} model with cointegration:
\begin{align*}
        d\log(S_t^d)&=\left(r-\frac{\sigma_{S^d}^2}{2}-\delta_t^d+b_d z_t\right) dt + \sigma_{S^d} dW_t^{S^d},\\
        d\delta_t^d&=\alpha_d \left(\mu_d-\delta_t^d\right)dt +\sigma_{\delta^d} dW_t^{\delta^d},
\end{align*}
for $d=1,\dots,D$, all the Brownian motions were correlated and 
\[
        z_t=\mu_Z+a_0 t + \sum_{d=1}^D a_d \log(S_t^d)
\]
was an error correction term related to cointegration. The authors then showed that, under certain assumptions, the dynamics of the process $z$ were stationary. They also showed that this model could obtain explicit formulas for futures prices. 
In 2015, Benth and Koekebakker~\cite{Benth2015b} were also interested in cointegration to price derivatives. They proposed, in turn, to generalize the Schwartz and Smith model for first the $D=2$ case, and then for the general multidimensional case: 
\[
        S_t^d=\exp(c_d X^0_t+ \sum_{n=1}^N A(d,n) X_t^{d,n}),
\]
with
\begin{align*}
        dX^0_t & =\mu dt + \sigma dW^0_t,\\
        dX_t^d &=K_d X_t^d dt + \sigma_d e_{p_d} dW_t^d,
\end{align*}
where the Brownian motions that appear were correlated. The vector $e_{p_d}$ was the last vector of the canonical basis of $\mathbb{R}^{p_d}$, and $K_d$ was a square matrix with $p_d$ rows and $p_d$ columns. The dynamics of $X^d$ were that of a CARMA process \cite{Benth2014a}. The cointegration came from the fact that the $X^0$ component was common to all prices. Some numerical results were given to analyze the performance of the model to value spread options.\par
The same authors also proposed a second model that they defined directly for futures prices. It was a Heath-Jarrow-Morton type model:
\[
        \frac{dF_t^d(T)}{F_t^d(T)}=\alpha_d(t,T) dt + c_d \sigma(t) dW_t + \sigma_d(T-t) dW_t^d, \quad d=1, \dots, D.
\]
This model was compatible with the previous model for a certain choice of $\sigma_d$ functions.\par 

\medskip
More recently, Farkas et al.~\cite{Farkas2017} propose a model for $D$ commodities which can take $1$ to $D-1$ cointegrating relations into account. The principle is to adopt a multidimensional two-factor model. The first factor is identified as the logarithm of the spot price (of dimension $D$) and
\[
\begin{split}
    \ln S_t & = \Lambda(t) + X^1_t, \\
    dX^1_t & = - K_1(X^1_t - X^2_t) dt + \Sigma^{1/2}_1 dW^1_t, \\
    dX^2_t & = (\mu - K_2 \Theta X^2_t)dt + \Sigma_{12}^{1/2} dW^1_t + \Sigma_{2}^{1/2} dW^2_t.
\end{split}
\]
Cointegration is present as soon as the rank of the $\Theta$ matrix is non-zero. The authors then consider the global variable $(X^1_t, X^2_t)$ which follows an Ornstein-\"Uhlenbeck process of dimension $2D$, whose conditional expectation and conditional covariance can be computed. After defining the probability change, they obtain an explicit formula for the futures prices. They then estimate the model for four electricity spot markets and price a spread option.

\bigskip
Correlations between Brownian motions are the simplest way to link several price dynamics. It allows, in general, to keep explicit formulas for the derivative prices. However, this technique is not rich enough to represent the fact that prices can be equal for several hours in two interconnected countries, or to obtain a stationary spread between two prices. Coupling or cointegration models are solutions to these problems. Such techniques have continued to develop in recent years, which shows that the needs they address are still relevant. Finally, it should be noted that the models proposed in this section, because of their simplicity, generally allow the expression of both spot and futures prices.
\section{Structural models} \label{section:modelesStructurels}
Structural models are a real specificity for the electricity market. This class of models has been able to develop in the particular case of electricity because, as Carmona and Coulon~\cite{Carmona2013} point out, the relation between the spot price and the fundamental elements that construct it, that is, the supply and demand curves, are better known and more observable than for other commodity markets. A\"id~\cite{Aid2015} calls a ``structural model'' any model that uses only observable variables as factors of uncertainty. In \cite{Carmona2013}, the idea of a structural model extends to using both observable variables and potentially unobservable factors. The common idea is that the spot price is written as a function of fundamental variables. We propose here to take these ideas and define the three main principles that, from our point of view, define a structural model:
\bit
\item the spot price is written as a function of fundamental variables ;
\item this function aims at reproducing the way the spot price is calculated, that is, by reproducing the principles of a balance between supply and demand; 
\item and the variables used are observable and their model is calibrated with observed data.
\eit
The construction function of the spot price from the fundamental variables can be more or less complex and can finely depict the reality of spot prices, while manipulating simple models on the fundamental variables. We can then see a first advantage of structural models: they allow us to finely model the characteristics of spot prices, such as price spikes (which is, in fact, the motivating element of most of the structural models proposed in the literature), while remaining simple models (generally Gaussian) on the fundamental variables, which avoids manipulating complex objects such as the L\'evy processes seen in the section~\ref{sec:pic}. Moreover, and as specified in \cite{Barlow2002} and \cite{Carmona2013}, structural models do not have the drawback of having extremely high mean-reverting effects as is the case when using L\'evy processes. Let us now specify that, as already stated in \cite{Carmona2013}, the function for constructing the spot price from the fundamental variables must be analytical (i.e. without having recourse to an optimization algorithm as is the case in the real market) to calculate the derivatives and, in particular, to calculate the futures prices. A second major advantage of structural models is that they create a more realistic link between fundamental variables and prices than by using a simple correlation; they also avoid the problem of estimating this correlation. This link is extremely important for many assets exposed to energy markets, such as thermal power plants that are sensitive to changes in the spread between electricity and fuel prices, renewable generation or electricity supply contracts that are sensitive to changes in electricity prices, and weather or consumption variables. In this class of structural modeling, some key concepts exist:
\bit
\item \textsl{merit order} is the principle that orders the means of production from the cheapest to the most expensive;
\item \textsl{marginality} is related to the most expensive means of production called on to satisfy the demand: one is the ``marginal cost'' that is the highest cost of production of the means used to satisfy the demand, and ``marginal technology'' or ``marginal fuel'' that is the most expensive technology or fuel which is used to satisfy the demand.
\eit
Table~\ref{tab:modele_structurel} summarizes the existing structural models for which the objective is to represent a single country. Despite the many differences in approach, one obvious similarity can be observed: demand is the fundamental variable common to all structural models, whatever the market studied. It is historically the most important variable to explain the behavior of electricity prices, which can be easily explained by the non-storable nature of electricity and therefore the need to satisfy the balance between supply and demand balance in real time.
\begin{table}[htb!]
	\centering
	\small
	\begin{tabular}{cccc}
		\toprule
		 Model  & \shortstack{Fundamental variables \\ considered} & \shortstack{Futures prices \\ computation} & Markets studied \\
		\midrule
		Eydeland and & Demand & \multirow{2}{*}{Analytical} & \multirow{2}{*}{US markets} \\
		Geman 1999~\cite{Eydeland1999} & Marginal fuel & & \\
		\midrule
		\multirow{2}{*}{Kosecki 1999~\cite{Kosecki1999}} & Demand & \multirow{2}{*}{Solution of PDE} & \multirow{2}{*}{US markets} \\
		 & Marginal fuel & & \\
		\midrule
		\multirow{2}{*}{Barlow 2002~\cite{Barlow2002}} & \multirow{2}{*}{Demand} & \multirow{2}{*}{Semi analytical} & Alberta \\
		 & & & California \\
		\midrule
		Burger & Demand & Analytical & \multirow{3}{*}{EEX} \\
		et al. 2004~\cite{Burger2004} & Maximal capacity & with & \\
		 & 2 unobservable factors & approximation & \\
		\midrule
		Kanamura and & \multirow{2}{*}{Demand} & \multirow{2}{*}{Semi analytical}  
		& \multirow{2}{*}{PJM} \\
		Ohashi 2007~\cite{Kanamura2007} & & & \\
		\midrule
		Pirrong and & Demand & \multirow{3}{*}{Solution of PDE} & \multirow{3}{*}{PJM} \\
		Jermakyan 2008~\cite{Pirrong2008} & Marginal fuel & & \\
		(Pirrong 2011)~\cite{Pirrong2011} & (Outage) & & \\
		\midrule
		Coulon and & Demand & Analytical & NEPOOL \\
		Howison 2009~\cite{Coulon2009} & Maximal capacity & case 1 fuel & PJM \\
		\midrule
		 A\"id et al. 2009~\cite{Aid2009} & Demand & \multirow{3}{*}{Analytical} & \multirow{3}{*}{France} \\ 
		 A\"id et al. 2013~\cite{Aid2013} & $K$ fuel costs & & \\
		  & $K$ capacities & & \\ \midrule
		Lyle and & Demand & \multirow{2}{*}{Analytical} & \multirow{2}{*}{Alberta} \\
		Elliot 2009~\cite{Lyle2009} & Baseload capacity & & \\ \midrule
		De Maere d'Aertrycke & Demand & \multirow{2}{*}{Solution of PDE} & \multirow{2}{*}{Germany} \\ 
		and Smeers 2010~\cite{DeMaere2010} & Marginal fuel & & \\ \midrule
		\multirow{4}{*}{Coulon et al. 2013~\cite{Coulon2013}} & Demand & \multirow{4}{*}{Analytical} & \multirow{4}{*}{ERCOT} \\
		 & Gas price & & \\
		 & 1 unobservable factor & & \\
		 & 1 state variable & & \\ \midrule
		\multirow{2}{*}{Carmona et al. 2013~\cite{Carmona2013}} & Demand & \multirow{2}{*}{Analytical} & \multirow{2}{*}{-} \\
		 & $K$ fuel costs & & \\ \midrule
		\multirow{3}{*}{Wagner et al. 2014~\cite{Wagner2014}} & Demand & \multirow{3}{*}{Monte Carlo} & \multirow{3}{*}{Germany} \\
		 & Wind production & & \\
		 & Solar production & & \\
		\bottomrule
	\end{tabular}
	\caption{\label{tab:modele_structurel} Summary of the characteristics of existing structural (uni-market) models }
	\label{tab:nb_factor}
\end{table}
\subsection{The initiating models}
We find the first modeling ideas that integrate fundamental variables in \cite{Eydeland1999} who curiously, propose a model on futures prices:
\[
f_t(T) = p_0 \phi(F^m_t(T), L_t(T))
\]
where $p_0$ is the ``base load'' futures price, $F^m_t(T)$ is the futures price of the ``marginal fuel'', and $L_t(T)$ is the demand forecast at time $T$ from time $t$. However, the authors use this model only to justify that in the particular case when function $\phi$ is exponential, this model is equivalent to considering stochastic volatility in futures prices. Another attempt is made in \cite{Kosecki1999} to define the price as the product of the marginal fuel price and a function of temperature. FPDs are then seen as the expectation of the average of daily spot prices over the delivery period. The ideas of this model are taken up by Pirrong and Jermakyan~\cite{Pirrong2008} who also consider the marginal fuel price and demand as random variables. They view the demand process as a controlled process that is bounded by the maximum production capacity. The authors describe in detail the probability change to obtain the dynamics under risk-neutral probability that allows them to obtain a partial differential equation for any derivative, and in particular for futures prices. However, some extensions of this model are given in chapter~8 of Pirrong~\cite{Pirrong2011} that notably introduce the  hazard of an outage, and also by De Maere d'Aertrycke and Smeers~\cite{DeMaere2010} with the consideration of several fuels and the introduction of a lower bound in the demand process.
Most papers, and notably \cite{Carmona2013} and \cite{Aid2015}, name the work of Barlow~\cite{Barlow2002} as the first structural model. It is indeed the first paper to describe an approach to reproduce supply and demand functions to reconstruct the spot price. As a result, this paper is the initiator of much of the subsequent work on structural models. The author defines the spot price as the price that equates the supply function and the demand function. He then very quickly makes the approximation that demand is inelastic, and this inelasticity is common to all structural models. He also makes the approximation that the supply function is constant over time to arrive at the relation:
\[
g(S_t) = D_t
\]
where $D_t$ is the demand level at time $t$, and $g$ is the supply function defined as:
\[
g(x) = a_0 - b_0x^\alpha
\]
in order to represent the fact that the total offer is limited. He then introduces a price cap so that the model is written as follows: 
\[
    S_t=\begin{cases} \left(\frac{a_0-D_t}{b_0}\right)^{1/\alpha} & \text{if }D_t<a_0-\varepsilon_0 b_0 \\ \varepsilon_0^{1/\alpha} & \text{if } D_t\geq a_0-\varepsilon_0 b_0 \end{cases}.
\]
The model chosen for demand is an Ornstein-\"Uhlenbeck process. Moreover, the futures prices can be calculated easily, even if there is no analytical formula.
Very quickly improvements appeared, notably in the representation of the supply curve, such as Kanamura and Ohashi~\cite{Kanamura2007} who propose a supply curve in the shape of a \textit{hockey stick} with several effects: (i) increasing the number of spikes and (ii) limiting the number of occurrences of price spikes to the allowed price cap. In addition, they consider seasonality on demand, which allows them to have price spikes at realistic times, especially in the summer in the case of the PJM market. 
\subsection{Exponential shape of the supply curve}
Burger et al.~\cite{Burger2004} introduce the concept of ``adjusted demand'' in the form of a ratio between demand and available production capacity. To the best of our knowledge, this is the first structural model that shows variability in production supply. The spot price model is then:
\[
 S_t = \exp \left(f(t, D_t/\bar{c}_t) + X^1_t + X^2_t \right)
\]
where $\bar{c}_t$ is the maximum (normalized) production capacity at time $t$. Moreover, the authors introduce unobservable factors $X^1$ and $X^2$ in order to ensure non-zero volatility in long-maturity futures prices. These factors make the model really usable in practice, unlike the structural models of this era which, with only short-term factors, do not allow volatile futures prices for long maturities, as illustrated in \cite{Aid2015} for the case of the Barlow~\cite{Barlow2002} model described above.
Pursuing the idea of random generation capacity, Cartea and Villaplana~\cite{Cartea2008} propose a simple expression for the spot price of electricity in $t$ as a function of the national demand $D_t$ and the capacity $C_t$ of a preponderant means of generation in the country: this capacity can be the total installed capacity of the country or the level of water reservoirs, if the country is highly dependent on hydro power. The model is:
\[
	S_t=\beta \exp(\gamma C_t+\alpha D_t),
\]
where $\alpha$ and $\beta$ are positive real numbers, and $\gamma$ is a negative real number. Without precisely reproducing the balance between supply and demand, this model seeks to reproduce a characteristic of prices: their growth with demand and their decrease with the available production capacity, all other things being equal. Demand and production capacity are modeled in a simple way by independent processes that are decomposed into a sum of a deterministic seasonality function and an Ornstein-\"Uhlenbeck process with time-dependent volatility functions.  \par 

\medskip
In 2013, Coulon et al.~\cite{Coulon2013} propose to keep a similar dynamics for the spot price and demand by adding the marginal fuel cost. The authors focused on the Texas market (ERCOT) where the marginal fuel price is considered constant and equal to the gas price. The authors also introduced an unobservable factor and a state variable for jumps to arrive at the following model:
\[
S_t = G_t \exp \left\{ \alpha_{z} + \beta_{z} D_t + \gamma_{z} X_t \right\} 
\]
in which $X$ is an unobservable factor, and $z$ is a state variable that has two values that describes two regimes characterized by parameters $\alpha_z$, $\beta_z$, and $\gamma_z$. In 2015, F\"uss et al.~\cite{Fuss2015} propose a model that was similar to the previous two. It keeps similar dynamics to the Cartea and Villaplana~\cite{Cartea2008} model for demand and capacity (but with an explicit volatility function for demand and constant for capacity). The authors emphasized the importance of the marginal fuel price, especially in countries where it was often the same from one day to the next. They then proposed the following model: 
\[
	S_t=\beta G_t^\delta\exp(\gamma C_t+\alpha D_t),
\]
in which $G_t$ was the marginal fuel price in $t$ and where the constraints on the parameters were the same as for Cartea and Villaplana~\cite{Cartea2008} after adding that $\delta>0$. The logarithm of the fuel price was modeled, again, by the sum of a deterministic function and an Ornstein-\"Uhlenbeck process. 
\\
The three papers~\cite{Cartea2008, Coulon2013, Fuss2015} perform the calculation of futures prices from their spot price model. The daily futures prices are given by the expectation of the spot price under a risk-neutral probability $\Q$, and then the FPDs are obtained using the relation~\eqref{eq_relation_produit_forward}. For Cartea and Villaplana~\cite{Cartea2008}, the choice of $\Q$ was made by introducing risk premium coefficients in the demand and capacity dynamics. Theoretical futures prices are calculated from these coefficients which were estimated by minimizing the difference between theoretical prices and market quotations. For Coulon et al.~\cite{Coulon2013} and F\"uss et al.~\cite{Fuss2015}, the approach was similar and also resulted in an estimation of the parameters of the measurement change by performing a difference minimization between theoretical and quoted FPDs.\par 

\medskip
The exponential form is again found in \cite{Lyle2009} where the spot price is defined as:
\[
S_t = \frac{1}{c} \left( \exp \left\{ - \frac{1}{b} \left(a C^b_t - D_t \right)\right\} - \xi \right)
\]
where $a$, $b$, $c$, and $\xi$ are real positive constants.  
However, the approach is very different from the previous two papers because the authors considered, as fundamental variables, the electricity demand and  a``baseload'' production capacity $C^b_t$ in $t$ that did not vary with price. Moreover, the way in which this exponential relation is constructed is based on a function that represented the top of the supply curve, that is, the ``spike'' part of production. The authors then found closed formulas for futures prices in two cases of modeling $C^b$: a classical Ornstein-\"Uhlenbeck model and a Markov chain model.
\subsection{Refinement of the supply curve: a merit-order principle}
The first representations of the supply curve in the form of a stacking function began to appear in 2009. Coulon et al.~\cite{Coulon2009} took up the initial idea of Barlow~\cite{Barlow2002} to represent the spot price in $t$ as the value of the supply function evaluated at the demand level $D_t$. Assuming that demand never exceeded the maximum production capacity, they interpreted the normalized supply function as a distribution function, and the spot price as a quantile of the corresponding distribution. They modeled the distribution with a mixture of laws and estimated with historical data on market orders. The interesting results of this work were that the different laws that composed the mixture identify with the different fuels that made up the market. In the case of a single fuel, the authors showed that the proposed model generalized that of Pirrong and Jermakyan~\cite{Pirrong2008}. The calculation of futures prices was, for the simplest cases, analytical. 
In \cite{Aid2009}, the first idea was again to link the supply function to the real market offers. The proposed representation was very simple, the supply function was a piecewise constant function that led to the following definition for the spot price at date $t$:
\[
S_t = \sum_{k=1}^K h^k S^k_t \unbb_{\sum_{j=1}^k C^k_t \le D_t < \sum_{j=1}^{k+1} C^k_t}
\]
in which $K$ is the number of generation technologies, $(C^k_t)_{k=1,\dots,K}$ are the generation capacities ordered according to their production cost, and $S^k_t$ and $h^k$ are respectively the fuel price and the heat rate of technology $k$. This approach actually describes the electricity price as the marginal cost at each date which makes the approximation that all market offers are made at the corresponding production cost. This simple approach has the advantage of giving a comprehensible form of electricity futures prices. Indeed, assuming independence between fuel prices and the fundamental variables of demand and capacity, and considering the appropriate probability change as explained in \cite{Aid2009}, we obtain the simple formula for futures prices at date $t$:
\[
F_t(T) = \sum_{k=1}^K F^k_t(T) \mathbb{P} \left(\sum_{j=1}^k C^k_T \le D_T < \sum_{j=1}^{k+1} C^k_T |\Fc_t \right)
\]
in which $F^k_t(T)$ is the futures price of fuel $k$, $\mathbb{P}$ is the actual probability, and $\Fc_t$ is the set of information available at date $t$. In other words, the futures price of electricity is an average of the futures prices of the fuels that is weighted by the marginality rate of the system (i.e., the marginality frequency of each technology). This is, to our knowledge, the first structural model that focuses on understanding the relation induced by the structural model on futures prices and is contrary to the previous models whose primary motivation was to properly represent spot prices. A classic criticism of this model is that it assumes a fixed merit order between fuels, which is not necessarily realistic. Indeed, there may be changes in fuel prices that modify the competitiveness of the technologies between them. It turns out that it is not difficult to incorporate possible permutations into the merit order, as explained, for example, in \cite{Alasseur2018}. On the other hand, this addition significantly increases the numerical cost for computing futures prices. 
An extension of this model is made in \cite{Aid2013} to try to introduce price spikes through a scarcity function:
\[
S_t = g\left(\overline{C}_t - D_t\right) \sum_{k=1}^K h^k S^k_t \unbb_{\sum_{j=1}^k C^k_t \le D_t < \sum_{j=1}^{k+1} C^k_t}
\]
where $\overline{C}_t = \sum_{k=1}^K C_t$ is the total capacity available at time $t$, and $g$ is a scarcity function defined as:
\[
g(x) = \min \left(M, \frac{\gamma}{x^\nu} \right) \unbb_{x >0} + M \unbb_{x < 0}
\]
where $\gamma$, $M$ and $\nu$ are positive constants. This function makes it possible to quickly increase the spot prices when the residual capacity $\overline{C}_t - D_t$ becomes very low (i.e., when the market becomes tense, the risk of default increases). The calculation of futures prices is then more complex, but can be obtained by a quasi-explicit formula and, under the same assumptions as in \cite{Aid2009}, the futures prices of electricity are expressed as a function of the futures prices of fuels. \par 
\medskip
The relations between electricity and the futures prices of fuel can be found in papers from the same period, such as in \cite{Carmona2013}, where the authors took a similar approach to that of A\"id et al.~\cite{Aid2009}, but with a richer supply function (a sum of exponential functions rather than a piecewise constant function) and a different way of incorporating price spikes as well as negative prices. The formulas obtained for futures prices become more complex and an example is given only in the 2-fuel case. The models of A\"id et al.~\cite{Aid2013} and Carmona et al.~\cite{Carmona2013} were, from our point of view, the most successful models in the class of structural models and were the ones that developed the most results in terms of derivatives valuation and optimal hedging strategies. Moreover, these models have been studied by other authors, in \cite{Feron2015} and \cite{Harms2019}, in terms of calibration performance and the sensitivity of the valuations to model parameters. Calibration on observed FPDs was also studied in \cite{Coulon2013} and constituted an essential element in the use of structural models for valuation. In general, the idea of calibration consists in introducing a risk premium to the processes of the fundamental (and unobservable) variables that make up the model. To the best of our knowledge, only Aid et al.~\cite{Aid2013} studied the probability change through hedging arguments in the classical framework of F\"ollmer and Schweizer~\cite{Follmer91}. 
A more recent paper by Moazeni et al.~\cite{Moazeni2016} further refines the supply curve by modeling the whole set of generation plants available in the PJM market. They consider a non-parametric modeling of the market supply. This original approach presents good results for spot price reconstruction. On the other hand, the complexity of the supply function makes it difficult to calculate futures prices because of the need to calculate a triple integral. To our knowledge, this is the only structural model that attempts to represent a zonal price.
Even if many papers focus on obtaining analytical or semi-analytical formulas for futures prices, the dynamics of long maturity futures products can have very little volatility and even be constant. This is explained by the very nature of structural models where futures prices depend on a future view of fundamental variables such as demand, capacity, etc. For long maturities, these future visions are not volatile (the vision of electricity demand in five years does not change every day). In this respect, we can cite an original contribution by Hinderks et al.~\cite{Hinderks2020a} who propose a hybrid model between the structural model and the HJM model. The structural model provides a future, structural view of the marginal cost of the system, and the HJM model represents the uncertainty arising from trading in futures prices. The model can be written as follows at date $t$:
\beqx
f_t(T) = X_t(T) \Ebb \left[CM_T | \Fc_t \right]
\eeqx
where $CM_t$ is the marginal cost whose expectation is obtained from a structural model, and $X_t(T)$ is independent of $CM_t$ and driven by a two-factor HJM model described by \eqref{eq:modele_2F} with drift zero and an initial value one for all $T \ge t$. This last constraint allows us to recover the martingale property of the futures prices.
\subsection{The case of interconnected markets}
The ideas of the structural approach have also been used to model more than one electricity price in interconnected markets. The structural approach aims to represent, in this case, the phenomena of interconnection saturation or price convergence. The first idea comes from Kiesel and Kustermann~\cite{Kiesel2016} who, in 2016, modeled interconnection saturation events between two countries to derive spot prices in the two markets. Specifically, at time $t$, they considered a single fuel and two price functions that depended on the price $G_t$ of the fuel, the demands $D_t^1$ and $D_t^2$ in each of the two interconnected markets, and the optimal energy transfer $E_t$ in the interconnections:
\[
\begin{array}{lcl}
    S^1_t & = & G_t e^{a_1 + b_1(D^1_t - E_t)} + c, \\
    S^2_t & = & G_t e^{a_2 + b_2(D^2_t + E_t)} + c.
\end{array}
\]
The authors then determined three disjoint event regions: (i) the case where the interconnection was saturated on the import side $E_t = E^{\max}_t$, (ii) the case where the interconnection was saturated on the export side $E_t = E^{\min}_t$, and (iii) the case where prices converged. They then used these three regions to break down the calculation of derivatives in a simple way, and in particular for futures prices. 
Similar modeling is in F\"uss et al.~\cite{Fuss2017} who extend their one-market model of \cite{Fuss2015} and allow different fuel prices in the two markets:
\[\begin{array}{lcl}
    S^1_t & = & \alpha_1 G_t^{\delta_1} e^{ b_1 (D^1_t - c_1  E_t)}, \\
    S^2_t & = & \alpha_2 G_t^{\delta_2} e^{ b_2 (D^2_t - c_2  E_t)}.
\end{array}
\]
Again, futures prices can be computed explicitly. 
An extension to the multi-fuel case is made by Alasseur and F\'eron~\cite{Alasseur2018}:
\[\begin{array}{lcl}
        S^1_t &= \left(\sum_{k=1}^{K_1}G^{1,k}_t \indiq_{D^1_t+E_t\in I^{1,k}_t}\right)e^{\alpha_1+\beta_1(\overline{C}^1_t-(D^1_t+E_t))},\\
         S^2_t &= \left(\sum_{k=1}^{K_2}G^{2,k}_t \indiq_{D^2_t+E_t\in I^{2,k}_t}\right)e^{\alpha_2+\beta_2(\overline{C}^2_t-(D^2_t+E_t))},\\
\end{array}
\]
where for each market $m=1,2$ and at time $t$; $K_m$ is the number of fuels, $G^{m,k}_t$ is the $k$\textsuperscript{th} cheapest fuel, and $I^{m,k}$ the marginality interval of fuel $k$. This is, to our knowledge, the most complex structural model to represent two interconnected countries; and it allows keeping semi-explicit formulas for futures prices as well as some important derivatives like spread options. These calculations are possible by taking up the ideas of Kiesel and Kustermann~\cite{Kiesel2016} of partitioning the space into disjoint regions according to the situations at the interconnections and the marginalities of each country. However, the formulas are quasi-explicit and require the calculation of the distribution function of a multivariate Gaussian distribution.

\bigskip
Structural models, which are very specific to electricity markets, have been very successful over the last 20 years. Their approach makes it possible to portray the links between fundamental variables and fuel prices on the one hand, and electricity prices on the other, without relying on the estimation of a correlation parameter in a reduced-form model. In fact, the modeling effort only focuses on the uncertainties in the balance between supply and demand. In recent years, structural models have developed considerably, as in the multi-country case treated by Alasseur and F\'eron~\cite{Alasseur2018}. However, the calculation of derivative prices can be complex, even for futures prices alone as in the major contributions of A\"id et al.~\cite{Aid2013} and Carmona et al.~\cite{Carmona2013} in using structural models for risk management purposes.
\section{Conclusion and view on the future}

\label{sec:conclusion}
Throughout this review, we have attempted to group research papers on electricity price modeling for valuation and financial risk management into different categories. Our categories are related to how the models represent the electricity price risk. Thus, we started by presenting Gaussian models based on interest rate models. These models have an undeniable advantage: they are simple to estimate, allow for the quick calculation of futures prices and conventional option prices, and enable hedging actions to be taken. This simplicity explains the success of this class of models. However, they do not reproduce certain characteristics of electricity prices such as spikes and extreme values in general.
\par
We then presented the natural extensions of Gaussian factor models that we initially introduced to depict price spikes. The use of L\'evy processes has allowed many researchers to enrich the dynamics of spot prices to represent these spikes, but at the cost of increasing the complexity of the derivatives calculation. Poisson processes and NIG processes are the most developed models in the literature, as they limit this complexity and provide quasi-analytical formulas for futures prices. The introduction of market incompleteness by these models also makes hedging more delicate, although solutions have been proposed by Goutte et al.~\cite{Goutte2013, Goutte2014} for example with the choice of a risk criterion to minimize. To represent price spikes, other researchers have proposed regime-switching approaches, but only one paper performed futures prices computations \cite{Kholodnyi2005}. 
\par
Another classical extension of Gaussian factor models has been the addition of stochastic volatility to better represent price distributions and option prices. This addition has led to the introduction of the more complex class of L\'evy semi-stationary (LSS) processes, which encompasses many models and provides a comprehensive theoretical framework. Again, the market incompleteness induced by these models generally increases the difficulty of computing options as well as determining optimal hedging strategies. \par
The consideration of multidimensional models is very useful in practice, given the natural exposure of energy assets to several markets. In this context, the dependency structure between the different prices modeled is very important. The first approaches consisted of accounting for simple correlations, then more refined dependencies with coupling variables or cointegration concepts. \par
Finally, we focused on structural models. These models have removed the randomness in electricity prices and replaced it with a fundamental view that reproduces the balance between supply and demand. This approach is promising for the representation of spot prices but makes risk management complex. However, it has benefited from thorough treatments, for example by A\"id et al.~\cite{Aid2013}. The calibration of structural models is also complex, since all the uncertainties used for the fundamental representation must be the subject of a modeling and estimation effort.\par 

\medskip
This review, together with the papers published in the last few years, allows us to draw some perspectives on what we believe to be the near future of electricity price modeling. CARMA-type processes (see, e.g., Brockwell~\cite{Brockwell2001}) and polynomial processes (developed notably by Ware~\cite{Ware2019} or by Kleisinger-Yu et al.~\cite{Kleisinger2020}) are promising because they can relatively easily compute futures prices while improving the dynamics of spot prices. Moreover, a recent article of Benth~\cite{Benth2021} performs a thorough summary of previous works on polynomial processes, giving formulas for futures prices and for the pricing of exotic options as well as outlooks on issues to investigate in order to perform accurate polynomial series expansions for exotic payoffs. From a theoretical point of view, LSS processes or ambit fields offer interesting frameworks but still suffer from practical difficulties.\par

The weather variables will have an increasing impact on electricity prices in the near future. First, the expected increase of renewable energy which is, by nature, impacted by weather conditions (wind, temperature, sun...) will have a high impact on electricity prices and their volatility. But also, we can expect the weather conditions to have a more important impact on the conventional thermal power plants, due to more frequent extreme climatic events. There already exists some literature on modeling weather variables like wind speed, temperature, etc. (see, for example, \cite{Benth2012f} for an overview), due to the existence of weather derivatives. However, to our knowledge, there is very little work about the joint modeling of weather variables and the electricity prices. 
While the structural models seem hardly adapted to introduce these variables, joint stochastic models of weather variables and prices, following, for example, the idea of \cite{Veraart2016, Deschatre2017, Rowinska2020}, can be expected to develop.\par 
As interconnections between markets expand, models develop to reproduce previously ignored facts, such as the equality of prices in multiple countries over multiple time periods. The recent work of Gardini et al.~\cite{Gardini2020a, Gardini2020b} and Christensen and Benth~\cite{Christensen2020} are notable advances in incorporating the connected nature of markets into modeling, without using the complexity of structural models on which the most recent work can only consider two interconnected markets.\par

Finally, the combined interest in energy markets (especially short-term) and trading models leads to an interest in the proper representation of liquidity, or more broadly of market incompleteness, for example in order to develop optimal trading strategies. On this subject, one can consult the work of Kremer et al.~\cite{Kremer2020} or Hess~\cite{Hess2017b}. We can also expect to see intraday markets increasingly focus industry interest as they capture value in flexible assets. The multi-horizon modeling of Hinderks et al.~\cite{Hinderks2020a} demonstrates the opportunities and challenges in terms of depicting such prices.
\bibliographystyle{siam}
\bibliography{Biblio,BiblioHorsEnergie,BiblioEnergieNonModele}
\end{document}